\documentclass[twocolumn,floatfix,aps,prd,showpacs,footinbib,amsmath,amssymb,amsfonts,superscriptaddress]{revtex4}
 \pdfoutput=1
\usepackage{bm}
\usepackage{graphicx}
\usepackage{color}
\usepackage{dsfont}
\begin{document}
\newcommand{\average}[1]{\langle{#1}\rangle_{{\cal D}}}
\newcommand{\dd}{{\rm d}}
\newcommand{\etal}{{\em et al.}}
\newcommand{\RTR}{{\mathbb R}^3}
\newcommand{\STR}{{\mathbb S}^3}
\newcommand{\HTR}{{\mathbb H}^3}
\newcommand{\LSS}{{\mathrm{LSS}}}

\newcommand{\PS}{{\cal P}_\phi}
\newcommand{\ALM}[2]{a_{{#1}\,{#2}}}
\newcommand{\CLMLPMP}[4]{C_{{#1}\,{#2}}^{{#3}\,{#4}}}
\newcommand{\SK}[1]{\mbox{s}_{#1}}
\newcommand{\ee}{{\rm e}}
\newcommand{\YLM}[2]{Y_{#1}^{#2}}
\newcommand{\PLM}[2]{P_{#1}^{#2}}
\newcommand{\GNA}[2]{C_{#1}^{#2}}
\newcommand{\JAC}[3]{P_{#3}^{({#1}\,,\,{#2})}}

\newcommand{\YXKLM}[4]{{\cal Y}^{[{#1}]}_{{#2}\,{#3}\,{#4}}}
\newcommand{\RXKL}[3]{R^{[{#1}]}_{{#2}\,{#3}}}
\newcommand{\UPSTKS}[3]{\Upsilon^{[{#1}]}_{{#2}\,{#3}}}
\newcommand{\XITKSLM}[5]{\xi^{[{#1}]\,{#3}}_{{#2}\,{#4}\,{#5}}}
\newcommand{\DIR}[1]{\delta^{{\rm D}}({#1})}
\newcommand{\KRON}[2]{\delta_{{#1}\,{#2}}}

\title{Topology beyond the horizon: how far can it be probed?}

\author{Oph\'elia Fabre}
\email{fabre@iap.fr}
\affiliation{Institut d'Astrophysique de Paris,
             Universit\'e Pierre~\&~Marie Curie - Paris VI,
             CNRS-UMR 7095, 98 bis, Bd Arago, 75014 Paris, France.}

\affiliation{Sorbonne Universit\'es, Institut Lagrange de Paris, 98 bis bd Arago,
             75014 Paris, France.}

\affiliation{Observatoire de Lyon, Universit\'e Claude Bernard, Lyon 1, 
              9 avenue Charles Andr\'e, Saint-Genis Laval, F-69230, France, 
             CNRS-UMR 5574, Centre de Recherche Astrophysique de Lyon.}

\author{Simon Prunet}
\email{prunet@iap.fr}
\affiliation{Institut d'Astrophysique de Paris,
             Universit\'e Pierre~\&~Marie Curie - Paris VI,
             CNRS-UMR 7095, 98 bis, Bd Arago, 75014 Paris, France.}

\affiliation{Sorbonne Universit\'es, Institut Lagrange de Paris, 98 bis bd Arago,
             75014 Paris, France.}

\author{Jean-Philippe Uzan}
\email{uzan@iap.fr}

\affiliation{Institut d'Astrophysique de Paris,
             Universit\'e Pierre~\&~Marie Curie - Paris VI,
             CNRS-UMR 7095, 98 bis, Bd Arago, 75014 Paris, France.}

\affiliation{Sorbonne Universit\'es, Institut Lagrange de Paris, 98 bis bd Arago,
             75014 Paris, France.}

\affiliation{Institut Henri Poincar\'e, Universit\'e Pierre et Marie Curie, 11 rue Pierre et Marie Curie, 75005 Paris, France}

\begin{abstract}
The standard cosmological model does not determine the spatial topology of the universe. This article revisits the signature of a non-trivial topology on the properties of the cosmic microwave background anisotropies. We show that the correlation function of the coefficients of the expansion of the temperature and polarization anisotropies in spherical harmonics, encodes a topological signature that can be used to distinguish a multi-connected space from an infinite space on sizes larger than the last scattering surface. The effect of the instrumental noise and of a galactic cut are estimated. We thus establish boundaries for the size of the biggest torus distinguishable with temperature and polarization CMB data. We also describe the imprint of the spatial topology on the 3-point function and on non-Gaussianity.
\end{abstract}
\date{13 November 2013}
 \maketitle

\section{Introduction}\label{section1}

In the standard cosmological framework in which the universe is described on large scales by a smooth Friedmann-Lema\^{\i}tre space-time, the spatial sections can enjoy a locally Euclidean, spherical or hyperbolic geometry~\cite{pubook}. Whatever the spatial geometry, it is always possible to assume different topologies for space, \emph{i.e.} different boundary conditions, a property that remains undetermined by the Einstein field equations. The study of spatial topology and the possibility to constrain it observationally has attracted a large activity in the past decades; see Refs.~\cite{revue} for reviews.

From an observational point of view, constraining the shape and size of our universe requires to use data spanning the largest possible scales compared to the Hubble volume. Initial works focused mostly on large scale structures~\cite{cristallo} but these techniques were limited by many effects, such as the completeness of the catalogues, evolution effects, \emph{etc.} and limited in the range of scales that can be probed. The cosmic microwave background (CMB) anisotropies seem to be the most promising observational tool for that purpose, mostly because they probe the largest cosmological scales we can currently access, have comparatively limited systematic errors and, from a theoretical point of view, only require the use of linear perturbation theory, which allow one to implement topology very efficiently~\cite{topo-cmb2,topo-cmb1}.

When decomposed in spherical harmonics, the mutipoles $a_{\ell m}$ of CMB temperature anisotropies are random complex fields characterised by their correlation matrix
\begin{equation}
 C_{\ell m}^{\ell'm'}\equiv\langle a_{\ell m} a^*_{\ell' m'}\rangle.
\end{equation}

Spatial topology imprints mostly three types of signatures on the CMB~\cite{ru-cras}: 
\begin{itemize}
 \item{\it Angular power spectrum}. This is the central quantity used to infer the standard cosmological constraints, mostly because as long as local isotropy (and thus statistical isotropy) holds then the correlation matrix reduces to
 \begin{equation}\label{e.diag}
 C_{\ell m}^{\ell'm'}\propto C_\ell \delta_{\ell\ell'} \delta_{mm'}
\end{equation}
so that the angular power spectrum $C_\ell$ contains the whole information of the temperature fluctuations if they are distributed according to Gaussian statistics. Since the $C_\ell$ is obtained by averaging the correlation matrix, it loses much of the topological information. For many years however, most constraints on the topology relied on its behaviour. It was used in the early analysis with COBE data, mostly to constrain the size of a torus universe~\cite{cmb-early}. One has however to note that (1) the topological signature appears on large angular scale, where the cosmic variance is the largest and (2) it depends on assumptions on the initial power spectrum. It is usually assumed that the initial power spectrum is almost scale invariant (as predicted by standard inflation, which at the same time predicts that the universe shall be locally Euclidean and much larger than the observable universe). Topology however sets a new cosmological characteristic scale and there is a priori no reason that the scale invariance of the power spectrum holds~\cite{Pinit}. Nonetheless, the low quadrupole observed by COBE and WMAP was one of the driving motivation for topology, in particular of the Poincar\'{e} space~\cite{pds}, and more generally to argue for a ``well-proportioned'' universe~\cite{CMB-lens,well-prop}. The angular power spectrum is thus a good indicator but will not be decisive in proving the existence or absence of any topological structure.
\item{\it Pattern correlation.} The CMB has been emitted at the time of last scattering so that all observed CMB photons arise from a 2-sphere centered on us. In a non-trivial topology, the last scattering surface can warp around and self-intersect on circles, which means, from an observer's point of view, that there exist pairs of circles sharing the same temperature anisotropy pattern along them. This point is at the heart of the ``circles in the sky method''~\cite{circle-method} that allows one, in principle, to detect and reconstruct~\cite{circle-back} space topology if it appears on scales smaller than the last-scattering diameter. We also refer to Refs.~\cite{circle-critics} for some critics concerning this method.
\item{\it Correlation function violation of global isotropy}. As long as isotropy holds, the correlation matrix is diagonal, in the sense that $\langle a_{\ell m} a_{\ell' m'}\rangle \propto \delta_{\ell\ell'}\delta_{mm'}$. A non-trivial topological structure implies that global isotropy is broken, which should be imprinted on deviations of the correlation matrix from a diagonal matrix; see \emph{e.g.} Refs.~\cite{ru-cras,T-correl1} for early considerations. This property was used to constrain the torus topology~\cite{torus-kunz}. The knowledge of the shape of the correlation matrix allows one to design adapted estimators and can be tested by different techniques~\cite{stat-aniso}.
\end{itemize}
In conclusion, the angular power spectrum is a poor indicator, the circle method is independent of assumptions on local physics but restricted to scales smaller than the diameter of the last scattering surface, $D_{\mathrm{LSS}}$. The information contained in the correlation matrix can be used to probe topology on larger scales, at the expense of being restricted to a class of topologies.

At the time being, cosmological observations indicate that space is almost Euclidean, which sets constraints on topologies that are potentially detectable~\cite{almost-flat}. The circles in the sky method was used with WMAP data, mostly to set the constrain that the length of the shortest closed space-like geodesic that self-intersects at our location in the universe to 98.5\% of $D_{\mathrm{LSS}}$, \emph{i.e.} about 26~Gpc~\cite{circle-wmap}. Independent constraints have been obtained for lens spaces~\cite{CMB-lens,CMB-lens2}. These constraints may be improved with the use of polarisation~\cite{polarization-topology} as well as with higher resolution data as provided by the \emph{Planck} mission (see Ref.~\cite{Planck_topology} for the \emph{Planck} analysis on topology). Note also that all existing observations of the CMB in COBE, WMAP and \emph{Planck} data have drawn special attention to several possible anomalies and statistical deviations from the standard model, such as North-South asymmetries, cold spot, axis of evil~\cite{evil-etal.}, point toward a possible violation of statistical isotropy and are the reasons underlying for the search for a new cosmological model.

From a theoretical point of view, the theory of CMB anisotropies in a non-trivial topology seems under control. It relies heavily on the linearity of the perturbation equations and of the temperature-perturbation relation arising from the Boltzmann equation at linear order. As shown in Ref.~\cite{topo-cmb2}, one can work in Fourier space so that the key ingredient to implement topology is the spectrum of the Laplacian. We will follow this technique that is summarized in Section~\ref{section2}.

In this article, we want to investigate the power of the correlation matrix method by restricting our analysis to a class of models in order to determine the minimal size of a non-trivial topology that makes CMB predictions undistinguishable from those of a universe with trivial topology. That question will be addressed by using the Kullback-Leibler divergence and described in Section~\ref{section3}. Section~\ref{section4} focuses on the family of torus universes in order to discuss their detectability, especially in the context of experimental issues such as masking and noise. Then section~\ref{section5} takes into account CMB polarization. To finish, we explore in Appendix~\ref{section6} the signature of spatial topology on higher statistics and on non-Gaussianity. This work will show that the topology of a flat torus can in principle be detected on scales larger that the last scattering surface even if one takes into account mask effects and noises.

In this work, the Python package Healpy based on {\tt HEALPix}~\cite{healpix} was used for all CMB simulations.

\section{Implementing the topology}\label{section2}

\subsection{General considerations}

The topology of three-dimensional spaces of constant curvature has been extensively described and we refer to the reviews~\cite{revue} for an introduction. For the sake of clarity, we just define the main structures required for our purpose.

In standard relativistic cosmology, the universe is described by a Friedmann-Lema\^{\i}tre spacetime with locally isotropic and homogeneous spatial sections. In the case of a multiply connected universe, we visualize space as the quotient $X / \Gamma$ of a simply connected space $X$ (which is just a 3-sphere $\STR$, a
Euclidean space $\RTR$, or a hyperbolic space $\HTR$, depending on the curvature), $\Gamma$ being a discrete and fixed point free symmetry group of $X$. If $\Gamma$ is not fixed point free, there is a curvature singularity and General Relativity can no longer be applied. For example string theory can be used with non-classical topologies with fixed point, as in Refs.~\cite{ben2012searching,rathaus2013orbifold}.
This group $\Gamma$ is called the holonomy group  and its existence changes the boundary conditions on all the functions defined on the spatial sections, which subsequently need to be $\Gamma$-periodic. Hence, the topology leaves the local physics unchanged while modifying the boundary conditions on all the fields. Given a field $\phi({\bm x},t)$ living on $X$, one can construct a field $\overline\phi({\bm x},t)$ leaving on $X / \Gamma$ by projection as
\begin{equation}
 \overline\phi({\bm x},t)=\frac{1}{\vert\Gamma\vert}\sum_{g\in\Gamma}\phi(g({\bm x}),t)
\end{equation}
since then, for all $g$, $\bar\phi(g({\bm x}),t)=\bar\phi({\bm x},t)$. It follows that any $\Gamma$-periodic function of $L^2(X)$ (space of square-integrable functions lying in the simply-connected space $X$) can be identified to a function of $L^2(X/\Gamma)$.

The background space-time being spatially homogeneous and isotropic, its metric is of the Friedmann-Lema\^{\i}tre form
\begin{equation}
 \dd s^2 = -\dd t^2 +a^2(t)\left[\dd\chi^2 + f_K^2(\chi)\dd\Omega^2 \right]
\end{equation}
where the scale factor $a$ is a function of the cosmic time $t$ and where $f_K(\chi)=\left\lbrace\sinh(\sqrt{-K}\chi)/\sqrt{-K},\chi,\sin(\sqrt{K}\chi)/\sqrt{K}\right\rbrace$ respectively for the comoving space curvature, $K$, negative, null, and positive.

The classification of the topologies of three dimensional spaces of constant curvature depends on the geometry of the universal covering space. In this article, we focus on the Euclidean space $\RTR$, for which there exists 18 different topologies that can be split into 10 compact spaces (6 orientable and 4 non-orientable), 5 chimney spaces having only two compact directions (2 orientable and 3 non-orientable), 2 slab spaces having one compact direction (1 orientable and 1 non-orientable) and the Euclidean space. Their holonomy group is a finite subgroup of the isometry of the Euclidean space $G = \RTR \times {\rm SO} (3)$. Their structure and fundamental polyhedron are given explicitly in Ref.~\cite{topo-cmb1}.

In the standard cosmological framework, the properties of large scale structures can be understood using perturbation theory. At linear order, the perturbation equations reduce to partial differential equations involving time derivatives up to second order and spatial derivatives, that appear only through a Laplacian because of the local spatial homogeneity of the background space-time. It is thus convenient to solve these equations in Fourier space where they become ordinary differential equations.

The strategy to implement topology is then in principle simple (we refer to Refs.~\cite{topo-cmb2,topo-cmb1} for early developments of this approach). First we shall solve the cosmological perturbation equations as in the standard framework but only for the eigenmodes of the Laplacian that are compatible with the boundary conditions imposed by the topology. One technical step is thus the determination of the eigenfunctions and we shall determine them by developing on the basis of the natural eigenfunctions of the Laplacian of the universal covering space. Then, the CMB predictions can be inferred from the linearity of the Sachs-Wolfe formula.

\subsection{Eigenmodes of the Laplacian}

Let us consider the usual Helmholtz equation

\begin{eqnarray}
\quad (\Delta \ + \ k^2) \ \Upsilon \  = \ 0
\end{eqnarray}

Once the topology is fixed, we must first determine the eigenmodes $\Upsilon^{[\Gamma]}_{\bm k}({\bm x})$ and eigenvalues $k^2 - K$ of the Laplacian on $X / \Gamma$ through the generalized Helmholtz equation
\begin{equation}\label{Helmholtz1} 
  \Delta\Upsilon^{[\Gamma]}_{\bm k}({\bm x})= -(k^2 - K) \Upsilon^{[\Gamma]}_{\bm k}({\bm x}) ,
\end{equation}
${\bm k}$ indexes the set of eigenmodes. These eigenmodes must satisfy the periodicity conditions
\begin{equation}
  \Upsilon^{[\Gamma]}_{\bm k}\left(g({\bm x})\right) = \Upsilon^{[\Gamma]}_{\bm k}({\bm x})\qquad\forall {\bm x}\in X/\Gamma,\quad\forall g\in\Gamma .
\end{equation}
These modes, on which any function on $X / \Gamma$ can be expanded, respect by the above definition the boundary conditions imposed by the topology: they correspond precisely to the modes of $X$ that are invariant under the action of the holonomy group $\Gamma$ so that any linear combination of such modes will satisfy, by construction, the required boundary conditions.

In order to compute CMB anisotropies, one needs to determine both the eigenvalues and eigenfunctions. It has been shown that it is fruitful to expand the modes of $X / \Gamma$ on the basis ${\cal Y}_{k\ell m}^{[X]}$ of the eigenmodes of the universal covering space as
\begin{equation}\label{eq:1} 
  \Upsilon^{[\Gamma]}_{\bm k} = \sum_{\ell = 0} \sum_{m = -\ell}^{\ell}  \xi^{[\Gamma]}_{{\bm k};\ell m}{\cal Y}_{k\ell m}^{[X]} ,
\end{equation}
so that all the topological information is now encoded in the coefficients $\xi^{[\Gamma]}_{{\bm k};\ell m}$. The sum over $\ell$ runs from $0$ to infinity if the universal covering space is non-compact (\emph{i.e.}, hyperbolic or Euclidean). These coefficients have been computed for many topologies and in particular  for all the Euclidean topologies~\cite{topo-cmb1}, the infinite class of spherical lens and prism topologies~\cite{mode-spherical}, and they require to be performed numerically for hyperbolic spaces~\cite{mode-hyperbolic}.\\

As a working example, we focus on the example of a rectangular 3-torus of comoving size $(L_{1},L_{2},L_{3})$. This implies that the wave-vectors, \emph{i.e.} the eigenvalues of the Helmholtz equation, are given by
\begin{equation}\label{k-discret}
 {\bm k} = 2\pi\left( \frac{n_1}{L_{1}}{\bm e}_x + \frac{n_2}{L_{2}}{\bm e}_y  + \frac{n_3}{L_{3}}{\bm e}_z  \right), \qquad
 {\bm n}\in {\mathbb Z}^3
\end{equation}
with the notations ${\bm n}=(n_1,n_2,n_3)$ and $n=\sqrt{{\bm n}\cdot{\bm n}}$. We also introduce
\begin{equation}
 \hat{\bm n}  \equiv  {\bm n} / n.
\end{equation}
The magnitude of the wave-number is defined as usual by $k^2={\bm k}\cdot{\bm k}$ and $\hat{\bm k} \equiv {\bm k}/k$.

In order to determine the eigenfunctions, we start from the fact that for any mode ${\bm k}$, the eigenmodes of the Laplacian of the universal covering space in  Cartesian coordinates are simply given by 
\begin{equation}\label{PlanarWave}
   \Upsilon_{\bm k} ({\bm x}) = \hbox{e}^{i {\bm k} \cdot {\bm x}}.
\end{equation}
These modes are indeed not square integrable and are normalized as
\begin{equation}\label{norm1}
\int \Upsilon_{\bm k} ({\bm x}) \Upsilon_{{\bm k}'}^* ({\bm x})
            \frac{\dd^3 {\bm x}}{(2 \pi)^3}
 = \delta^{(3)}({\bm k} - {\bm k}'),
\end{equation}
$ \delta^{(3)}$ being the Dirac distribution. As can be seen from Eq.~(\ref{eq:1}), we need to know the eigenmodes of the Laplacian in spherical coordinates. For the Euclidean space, they can be decomposed as the product of a radial and an angular part as
\begin{equation}\label{SphericalWave}
{\cal Y}_{k \ell m} (\chi, \theta, \varphi) = \sqrt{\frac{2}{\pi}} \,
   (2 \pi)^{3 / 2} \, j_\ell (k \chi) \, Y_\ell^m (\theta, \varphi) ,
\end{equation}
where $(\chi, \theta, \varphi)$ are the usual spherical coordinates defined by
\begin{eqnarray} \label{SphericalCoordinates}
x & = & \chi \; \sin{\theta} \; \cos{\varphi} \nonumber \\
y & = & \chi \; \sin{\theta} \; \sin{\varphi} \nonumber \\
z & = & \chi \; \cos{\theta} .
\end{eqnarray}
The radial factor $j_\ell (k \chi)$ is a spherical Bessel function of index $\ell$, and the angular factor $Y_\ell^m (\theta, \varphi)$ is the standard spherical harmonic.  The mode ${\cal Y}_{k \ell m}$ is not square integrable and is normalized according to
\begin{equation}
\int {\cal Y}_{k \ell m} {\cal Y}_{k' \ell' m'}^*
            \frac{\chi^2 \dd \chi \dd \cos \theta \dd \varphi}{(2 \pi)^3}
 = \frac{1}{k^2} \delta^{(1)}(k - k') \delta_{\ell\ell'} \delta_{mm'} ,
\end{equation}
which is analogous to the normalization (\ref{norm1}) and which
determines the numerical coefficient $\sqrt{2 / \pi}$.

The coefficients $\xi^{[\Gamma]}_{{\bm k};\ell m}$ need to express the Cartesian eigenmodes in terms of the spherical eigenmodes.
Using Eqns 5.17.3.14 and 5.17.2.9 of Ref.~\cite{vmk}, we have
\begin{eqnarray}\label{Conversion}
\hbox{e}^{i {\bm k} \cdot {\bm x}} 
 & = & \sum_{\ell = 0}^\infty \;
            i^\ell \; j_\ell (k \, |{\bm x}|) \; (2\ell + 1) \; 
            P_\ell (\cos{\theta_{{\bm k}, {\bm x}}}) \nonumber \\
 & = & \sum_{\ell = 0}^\infty \;
            i^\ell \; j_\ell (k \, |{\bm x}|) \;
            \left(4 \pi \sum_{m = - \ell}^{\ell}
                             Y_\ell^m (\hat{\bm x}) Y_\ell^{m*} (\hat{\bm k})
            \right) \nonumber \\
 & = & \sum_{\ell = 0}^\infty \,
            \sum_{m = - \ell}^{\ell} \;
                 i^\ell \; Y_\ell^{m*} (\hat{\bm k}) \;
                 \left[ \; 4 \pi \; j_\ell (k \, |{\bm x}|) \;
                        Y_\ell^m (\hat{\bm x}) \; \right] \nonumber \\
 & = & \sum_{\ell = 0}^\infty \,
            \sum_{m = - \ell}^{\ell} \;
                 \left( \; i^\ell \; Y_\ell^{m*} (\hat{\bm k}) \right) 
                 {\cal Y}_{k l m} ({\bm x}) ,
\end{eqnarray}
where $\hat{\bm x} \equiv {\bm x} / |{\bm x}|$. Comparing with Eq.~(\ref{eq:1}) it gives
\begin{equation}\label{xi-torus}
\xi^{[\Gamma]}_{{\bm k} \ell m} = i^\ell \; Y_\ell^{m*} ({\bf \hat k}).
\end{equation}

\subsection{Implementations and tests}

\subsubsection{CMB primer}

The CMB is observed as a black-body radiation with temperature $T_0 = 2.7255 \pm 0.0006 \,{\rm K}$~\cite{fixsen}, almost independently of the direction.  After accounting for the peculiar motion of the Sun and Earth, the CMB has remaining temperature fluctuations of order $\delta T / T_0 \sim 10^{- 5}$ that are usually decomposed in terms of spherical harmonics as
\begin{equation}\label{dttalm}
 \frac{\delta T}{T_0} (\theta, \varphi)  = \sum_{\ell = 0}^\infty \sum_{m = - \ell}^\ell
    a_{\ell m} Y_{\ell m} (\theta, \varphi) .
\end{equation}
This relation can be inverted by using the orthonormality of the spherical harmonics to get
\begin{equation} \label{p4}
  a_{\ell m}  = \int \frac{\delta T}{T_0}Y_{\ell m}^* \sin \theta \dd \theta \, \dd \varphi .
\end{equation}
The coefficients $a_{\ell m}$ obviously satisfy $a_{\ell -m}=(-1)^m a_{\ell m}^*$. The angular correlation function of these temperature anisotropies is observed on a $2$-sphere around us and can be decomposed on a basis of the Legendre polynomials $P_\ell$ as
\begin{eqnarray} \label{dTT}
C^{\rm obs} (\theta_{12})  &=& 
\left< \frac{\delta T}{T_0}(\hat \gamma_1)\frac{\delta T}{T_0}(\hat \gamma_2) \right>_{\hat \gamma_1 . \hat \gamma_2 = \cos \theta_{12}}\nonumber\\
 &=& \frac{1}{4 \pi} \sum_\ell (2 \ell + 1) C_\ell^{\rm obs} P_\ell(\cos \theta_{12}),
\end{eqnarray}
where the brackets stand for an average on the sky, \emph{i.e.}, on all pairs of directions $(\hat \gamma_1, \hat \gamma_2)$ subtending an
angle $\theta_{12}$. The coefficients $C_\ell^{\rm obs}$ of the development of $C^{\rm obs} (\theta_{12})$ on the Legendre polynomials
are thus given by
\begin{equation}
C_\ell^{\rm obs}  = \frac{1}{2 \ell + 1}   \sum_{m = - \ell}^\ell
   \left \langle a_{\ell m}^{\rm obs} a_{\ell m}^{\rm obs} {}^* \right\rangle .
\end{equation}
These $C_\ell^{\rm obs} $ can be seen as estimators of the variance of the $a_{\ell m}$  and represent the rotationally invariant angular power spectrum. They have therefore to be compared to the values $C_\ell$ predicted by a given cosmological model, which is specified by (i) a model of structure formation which fixes the initial conditions for the perturbation (\emph{e.g.} inflation, topological defects, etc), (ii) the geometry and matter content of the universe (via the cosmological parameters) and (iii) the topology of the universe.

In the particular case of a Euclidean space that we are considering here, the temperature fluctuation in a given direction of the sky can be related to (i) the eigenmodes $\exp (i {\bm k} \cdot {\bm x})$ of the Laplacian by a linear convolution operator $O_k^{[\RTR]}(e^{i {\bm k} \cdot {\bm x}})$ depending on the modulus $k$, the universal cover (here, $\RTR$) and the cosmological parameters, and (ii) a 3-dimensional variable $\hat e_{\bm k}$ related to the initial conditions,
\begin{equation}\label{p44b}
 \frac{\delta T}{T_0}(\theta, \varphi) = \int \frac{\dd^3 {\bm k}}{(2 \pi)^{3 / 2}}
        O_k^{[\RTR]} \left(\ee^{i{\bm k}\cdot{\bm x}} \right)
        \sqrt{\PS (k)} \hat e_{\bm k} ,
\end{equation}
where $\PS (k)$ is the gravitational potential initial power spectrum, normalized so that $\PS (k) \propto k^{- 3}$ for a Harrison-Zel'dovich spectrum. The detail of the transfer function encoded in the operator $O_k^{[\RTR]}(e^{i {\bm k} \cdot {\bm x}})$ is described in Eq.~(\ref{swdopisw}). In most inflationary models the random variable $\hat e_{\bm k}$ describes a Gaussian random field and satisfies
\begin{equation}
  \left<\hat e_{\bm k} \hat e_{{\bm k}'}^* \right> = \delta^{(3)}({\bm k} - {\bm k}')
\end{equation}
with $\hat e_{-{\bm k}}=\hat e_{{\bm k}}^*$. Decomposing the exponential as in Eq.~(\ref{Conversion}), one gets
\begin{equation}
\frac{\delta T}{T_0}(\theta, \varphi)  = 
       \sum_{\ell, m} i^\ell \int k^2 \dd k \sqrt{\PS (k)}
       O_k^{[\RTR]}\! \left(\YXKLM{\RTR}{k}{\ell}{m}\right) \hat e_{\ell m}(k) ,
\end{equation}
with
\begin{equation}
 \hat e_{\ell m}(k)
 \equiv \int \dd \Omega_{\bm k}
             \YLM{\ell}{m}{}^* (\theta_{\bm k}, \varphi_{\bm k}) \hat e_{\bm k}.
\end{equation}
This quantity is a 2-dimensional Gaussian random variable satisfying
$\left< \hat e_{\ell m}(k) \hat e_{\ell'm'}^*(k') \right> = \delta(k- k') \delta_{\ell\ell'} \delta_{mm'} / k^2$.  It follows that the
coefficients $\ALM{\ell}{m}$ take the general form
\begin{equation}\label{p4_2}
 a_{\ell m} = i^\ell \int k^2 \dd k \sqrt{\PS (k)}  G_\ell (k) \hat{e}_{\ell m} (k) ,
\end{equation}
with $G_\ell (k) = O_k^{[\RTR]} \left(\RXKL{\RTR}{k}{\ell}\right)$, and $\RXKL{\RTR}{k}{\ell}=\sqrt{2 \over \pi}j_\ell[k(\eta_0-\eta_{LSS})]$. The transfer function is well approximated by (see, \emph{e.g.}, Refs.~\cite{abott86,cmb4})
\begin{widetext}
\begin{eqnarray}\label{swdopisw}
G_\ell (k)
 & = &   j_\ell [k (\eta_0 - \eta_\LSS)]
         \left(  \frac{\delta T}{T_0} (k, \eta_\LSS)
               + \Phi (k, \eta_\LSS) + \Psi (k, \eta_\LSS)
         \right)
       + j'_\ell [k (\eta_0 - \eta_\LSS)] \frac{v_{\rm b}(k, \eta_\LSS)}{k}
\nonumber \\ & &
       + \int_{\eta_\LSS}^{\eta_0} j_\ell [k (\eta_0 - \eta)]
         \left(\dot \Phi (k, \eta) + \dot \Psi (k, \eta) \right) \; \dd \eta.
\end{eqnarray}
\end{widetext}
$\eta_\LSS$ and $\eta_0$ are the conformal times at last scattering and today, $\Phi$ and $\Psi$ are the two Bardeen potentials, and $v_{\rm b}$ is the velocity divergence of the baryons. The first term is the Sachs-Wolfe contribution, the second one the Doppler contribution and the last one the integrated Sachs-Wolfe contribution \cite{pubook}. As the topology is the study of large scales, \emph{i.e.} low multipoles, we will mainly use only the Sachs-Wolfe contribution, instead of the full transfer function, in our analysis.

\subsubsection{Implementing topology}

The topology does not affect local physics, so the equations describing the evolution of the cosmological perturbations are left unchanged. As a consequence, quantities such as the Bardeen potentials $\Phi$, $\Psi$, \emph{etc.}, are computed in the same way as in the standard case, and the operator $O_k^{[X]}$ is therefore the same. However, a change of topology translates into a change of the modes that can be excited. We thus need to decompose the perturbation on the basis of $\Upsilon_{\bm k}$ instead of ${\cal Y}_{k\ell m}$. 

Using Eq.~(\ref{eq:1}) and the fact that the convolution operator $O_k^{[X]}$ is linear, Eq.~(\ref{p44b}) now takes the
form 
\begin{equation} \label{p44t}
  \frac{\delta T}{T_0} (\theta, \varphi)  = \frac{(2\pi)^3}{V}
   \sum_{\bm k} O_k^{[X]} \left(\Upsilon^{[\Gamma]}_{\bm k}\right) \sqrt{\PS (k)} \hat{e}_{\bm k} ,
\end{equation}
where now $\hat{e}_{\bm k}$ is a 3-dimensional random variable which is related to the discrete mode ${\bm k}$. These random variables satisfy the normalization
\begin{equation}
 \left< \hat{e}_{\bm k} \hat{e}_{{\bm k}'}^* \right> = \frac{V}{(2 \pi)^3} \delta_{\bm k \bm k'}.
\end{equation}
By inserting the expansion of $\Upsilon_{\bm k}$ in terms of the covering space eigenmodes, as given by Eq.~(\ref{eq:1}), we obtain
\begin{equation} \label{p44tbis}
\frac{\delta T}{T_0}(\theta, \varphi)
 = \frac{(2 \pi)^3}{V} \sum_{k, s} \sum_{\ell, m}
                    \XITKSLM{\Gamma}{k}{s}{\ell}{m}
                     O_k^{[X]} \left(\YXKLM{X}{k}{\ell}{m}\right)
               \sqrt{\PS (k)} \hat{e}_{\bm k} .
\end{equation}
It follows that the $\ALM{\ell}{m}$, seen as random variables, are given
by
\begin{equation}\label{p44t2}
\ALM{\ell}{m}
 = \frac{(2 \pi)^3}{V}
   \sum_k \sqrt{\PS (k)} O_k^{[X]} \left (\RXKL{X}{k}{\ell} \right)
          \sum_s \XITKSLM{\Gamma}{k}{s}{\ell}{m} \hat{e}_{\bm k}.
\end{equation}
Note that the sum over $s$ is analogous to the sum over angles
defining the 2-dimensional random variable $\hat e_{\ell m}$ in
Eq.~(\ref{p44b}). Since the $a_{\ell m}$ are linear functions
of the initial 3-dimensional random variables, they are still
Gaussian distributed but they are not independent anymore (as
explained before, this is the consequence of the breakdown of
global isotropy and/or homogeneity). The correlation between the
coefficients $\ALM{\ell}{m}$ is given by
\begin{eqnarray}\label{corr_mat}
\left< \ALM{\ell}{m} \ALM{\ell'}{m'}^* \right>&
 = &\frac{(2 \pi)^3}{V}
   \sum_k \PS (k) O_k^{[X]} \left(\RXKL{X}{k}{\ell} \right)
                  O_k^{[X]} \left(\RXKL{X}{k}{\ell'} \right)\nonumber\\
                  && \sum_s \XITKSLM{\Gamma}{k}{s}{\ell}{m}
                 \XITKSLM{\Gamma}{k}{s \; *}{\ell'}{m'} .
\end{eqnarray}
Clearly these correlations can have non-zero off-diagonal terms,
reflecting the global anisotropy induced by the multi-connected
topology.  This means in
particular that for fixed $\ell$, the $\ALM{\ell}{m}$ might not have
the same variance, although they all follow Gaussian statistics as
long as the initial conditions do. This translates into an apparent
non-Gaussianity in the sense that the $C_\ell$ will not follow the
usual $\chi^2$ distribution.  Strictly speaking, this is not a
signature of non Gaussianity but of anisotropy.

Note also that the correlation matrix~(\ref{corr_mat}) is not rotation
invariant. It explicitly depends on the orientation of the
manifold with respect of the coordinate system. However, knowing how
the spherical harmonics transform under a rotation allows us to
compute the correlation matrix under any other orientation of the
coordinate system. To finish let us note that one can define the usual
$C_\ell$ coefficients in any topology by the formula
\begin{equation}
C_\ell \equiv \frac{1}{2\ell + 1} \sum_m \CLMLPMP{\ell}{m}{\ell}{m} ,
\end{equation}
which is easily shown to be rotationally invariant. The $C_\ell$ coefficients can be generalized for higher statistical orders. In Appendix~\ref{section6}, we present the computation of the equivalent coefficients $B_{\ell \ell \ell}$ for the 3-point function of a torus.

\subsubsection{Tests}

\begin{figure}[h!]
\centering
\includegraphics[width=0.7\columnwidth]{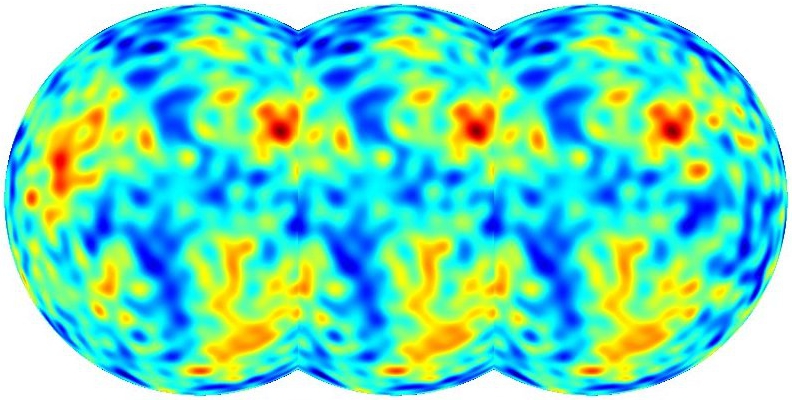}
\caption{This figure is composed of three identical spherical CMB temperature
    maps with a resolution $\ell_{\rm max}=50$ of the same cubic torus universe whose
    edge is of size  $L=R_{\rm LSS}$. The origin of the framework is the center of
    the middle sphere. The distance between the centers of two neighbouring
    spheres is equal to $R_{\rm LSS}$. Let us assume that we are living in the
    universe whose CMB temperature fluctuations are displayed by the middle
    sphere. Thanks to this example, we can illustrate that (i) there is a
    periodic pattern in the CMB due to an invariance of translation $\vec{t}$ along
    the $\hat{y}$ axis, with $||t||=R_{\rm LSS}$, and (ii) the intersection of the last
    scattering surface with itself in a circle pattern in the CMB and we
    thus have pairs of correlated circles; for example, in this figure, we
    can check the existence of two correlated circles whose centers are
    $C_1=(0,-R_{\rm LSS}/2,0)$ and $C_2=(0,R_{\rm LSS}/2,0)$ and whose radii are equal to
    $\sqrt{3}R_{\rm LSS}/2$.}
\label{fig1}
\end{figure}

\begin{figure}[h!]
\centering
 \includegraphics[width=\columnwidth]{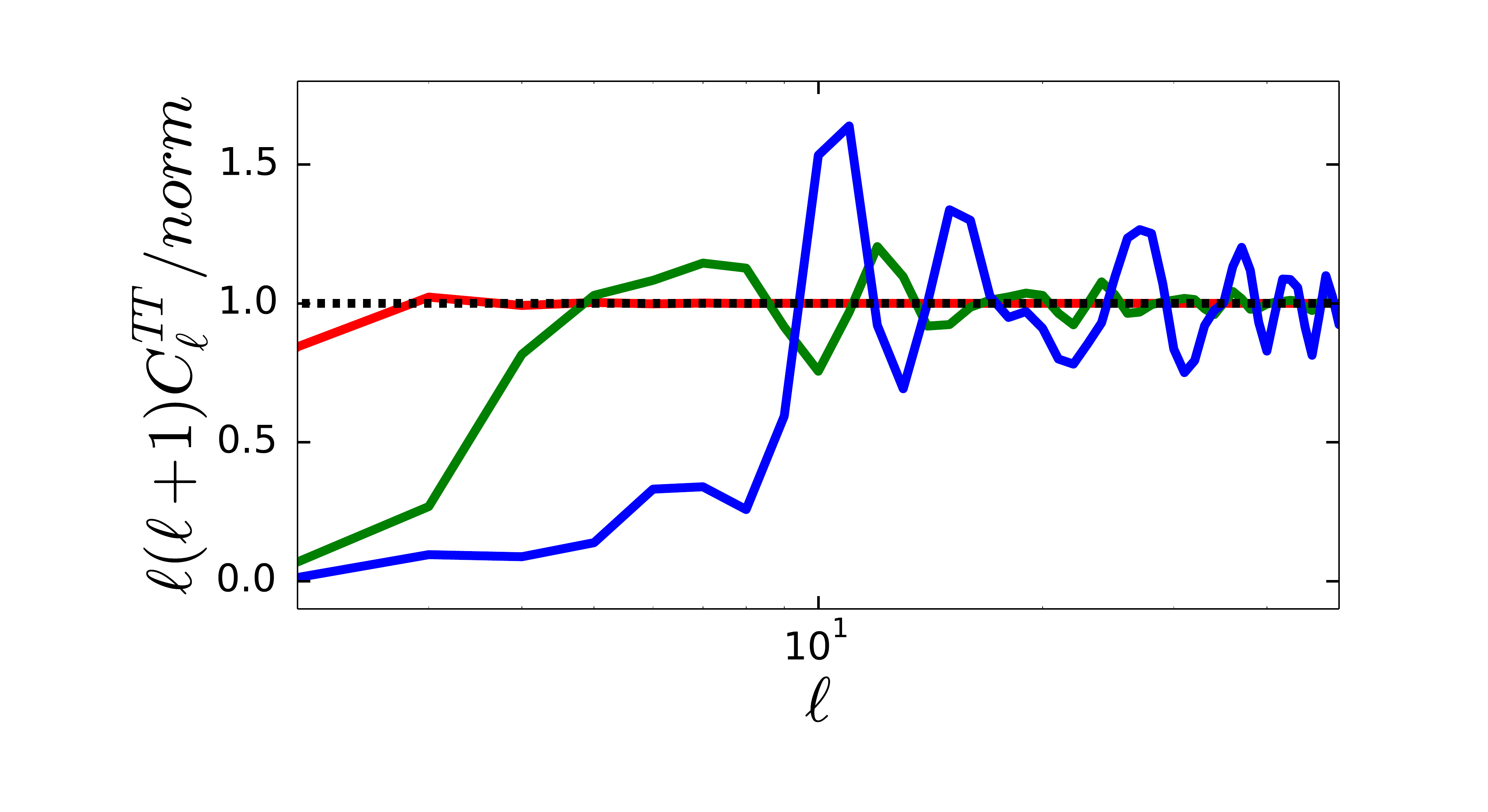}
\caption{Angular power spectrum $C_\ell$ for universes with the topology of a cubic torus with size $L=0.25D_{\mathrm{LSS}}$ (blue), $0.5D_{\mathrm{LSS}}$ (green) and 
$D_{\mathrm{LSS}}$ (red), compared with the angular power spectrum of the Euclidean space (black). The norm is taken to have the angular spectrum of the isotropic space at a plateau equal to one. All the computation were done by taking into account only the Sachs-Wolfe effect (justified for small $\ell$).}
\label{fig1b}
\end{figure}

The $\CLMLPMP{\ell}{m}{\ell}{m}$ generation code was implemented by picking the $\XITKSLM{\Gamma}{k}{s}{\ell}{m}$ defined in Eq.~(\ref{xi-torus}), then computing the discrete wave modes of the topology in an octile of ${\mathbb{N}}^3$, reversing the non-zero components in order to cover the whole space and summing these wave modes as described by Eq.~(\ref{corr_mat}). We validate the code by testing the existence of the ``circles-in-the-sky" properties and the behaviour of the angular power spectrum already discussed in the literature. The investigation of the existence of circles can only be pursued for sizes $L$ smaller than $D_{\mathrm{LSS}}$.  Our code reproduces with great success these CMB pattern properties as represented explicitly in Fig.~\ref{fig1}. We also compute the $C_\ell$ angular power spectrum for different sizes of cubic 3-tori as illustrated in Fig.~\ref{fig1b}. We got the same qualitative results than in Ref.~\cite{topo-cmb2}, concerning the damping of the curve at low $\ell$ and the great amplitude oscillations of the plot for large $\ell$.

\subsubsection{Correlation matrices}

We remind that it has been demonstrated in Ref.~\cite{topo-cmb2} that
\begin{eqnarray}\label{rel}
   \CLMLPMP{\ell}{m}{\ell'}{m'} \in {\mathbb R} 
\end{eqnarray}
and that
\begin{equation}\label{e.34}
\CLMLPMP{\ell}{m}{\ell'}{m'}
 = \frac{1}{4} \left[1 + (-1)^{m - m'} \right]
               \left[1 + (-1)^{\ell - \ell'} \right]
   \CLMLPMP{\ell}{m}{\ell'}{m'} ,
\end{equation}
so that $\CLMLPMP{\ell}{m}{\ell'}{m'} \neq 0$ if $m - m' \equiv 0 \,{\rm mod}\,{2}$ and $\ell - \ell' \equiv 0\, {\rm mod}\,{2}$. Furthermore,
\begin{equation} \label{toregen2}
\CLMLPMP{\ell}{m}{\ell'}{m'} = \CLMLPMP{\ell}{- m}{\ell'}{- m'} .
\end{equation}
These properties of the correlation matrix hold for any torus. 

In the particular case of a cubic torus, there exists an invariance under a $\pi / 2$-rotation about the $z$ axis, so if
$(n_1, n_2, n_3)$ corresponds to a wave-number then so does $(n_2,- n_1, n_3)$, and one has
\begin{equation}\label{torecub}
\CLMLPMP{\ell}{m}{\ell'}{m'} \neq 0
\quad \Rightarrow \quad
m - m' \equiv 0 \,{\rm mod}\,{4} .
\end{equation}

In the following, we considered the normalized correlation matrix defined by
\begin{equation}\label{normalizedC}
   C_{ss'}=\frac{\left<a_s a_{s'}^*\right>}{\sqrt{\left<a_s a_{s}^*\right>\left<a_{s'} a_{s'}^*\right>}},
\end{equation}
that is the correlation matrix normalized to the angular power spectrum where we have used the notation
\begin{equation}
 s\equiv\ell(\ell+1)+m,\qquad -\ell\leq m\leq\ell,
\end{equation}
so that $s$ is strictly increasing.

\begin{figure*}[htb]\label{fig2}
\centering
\includegraphics[width=5.8cm]{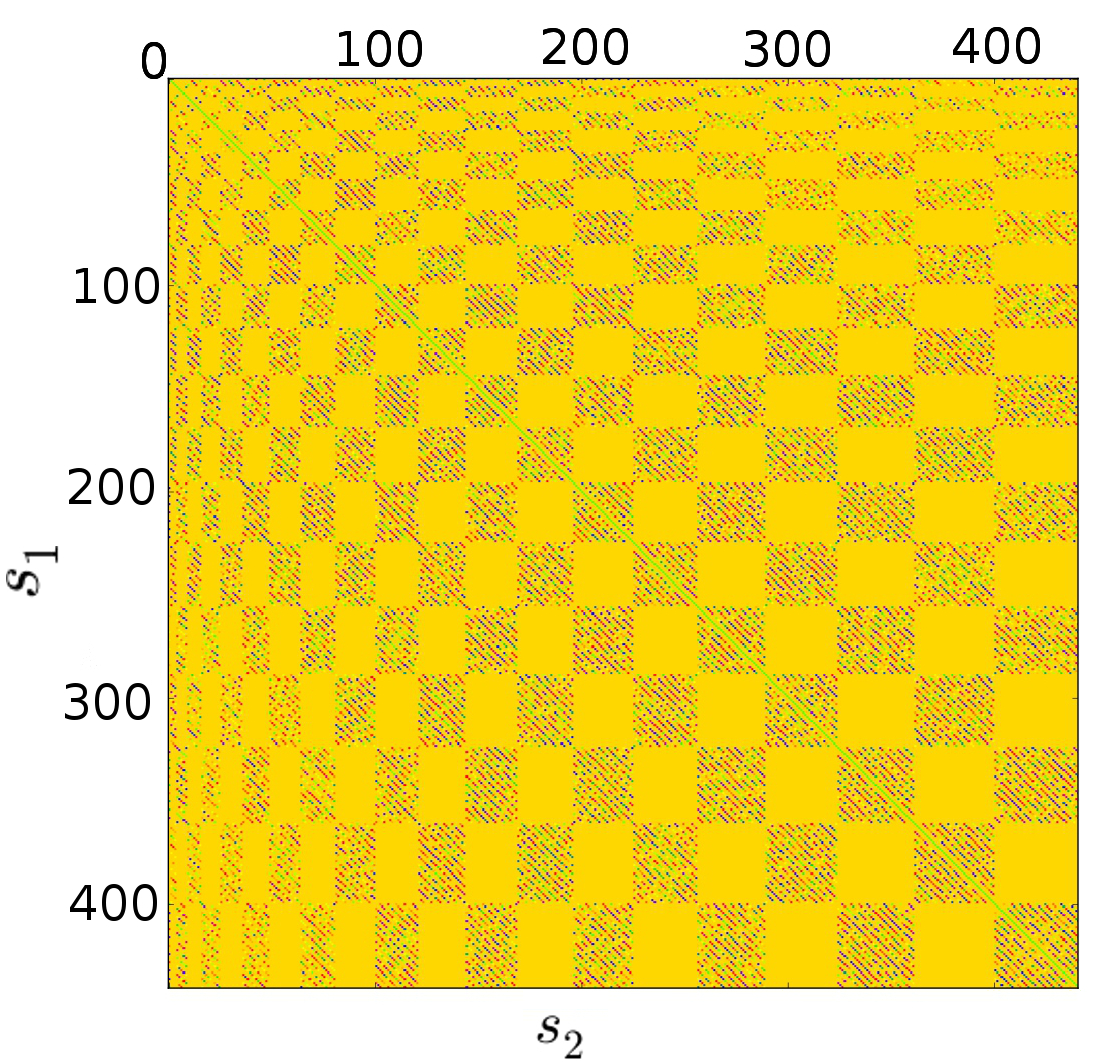}\includegraphics[width=5.8cm]{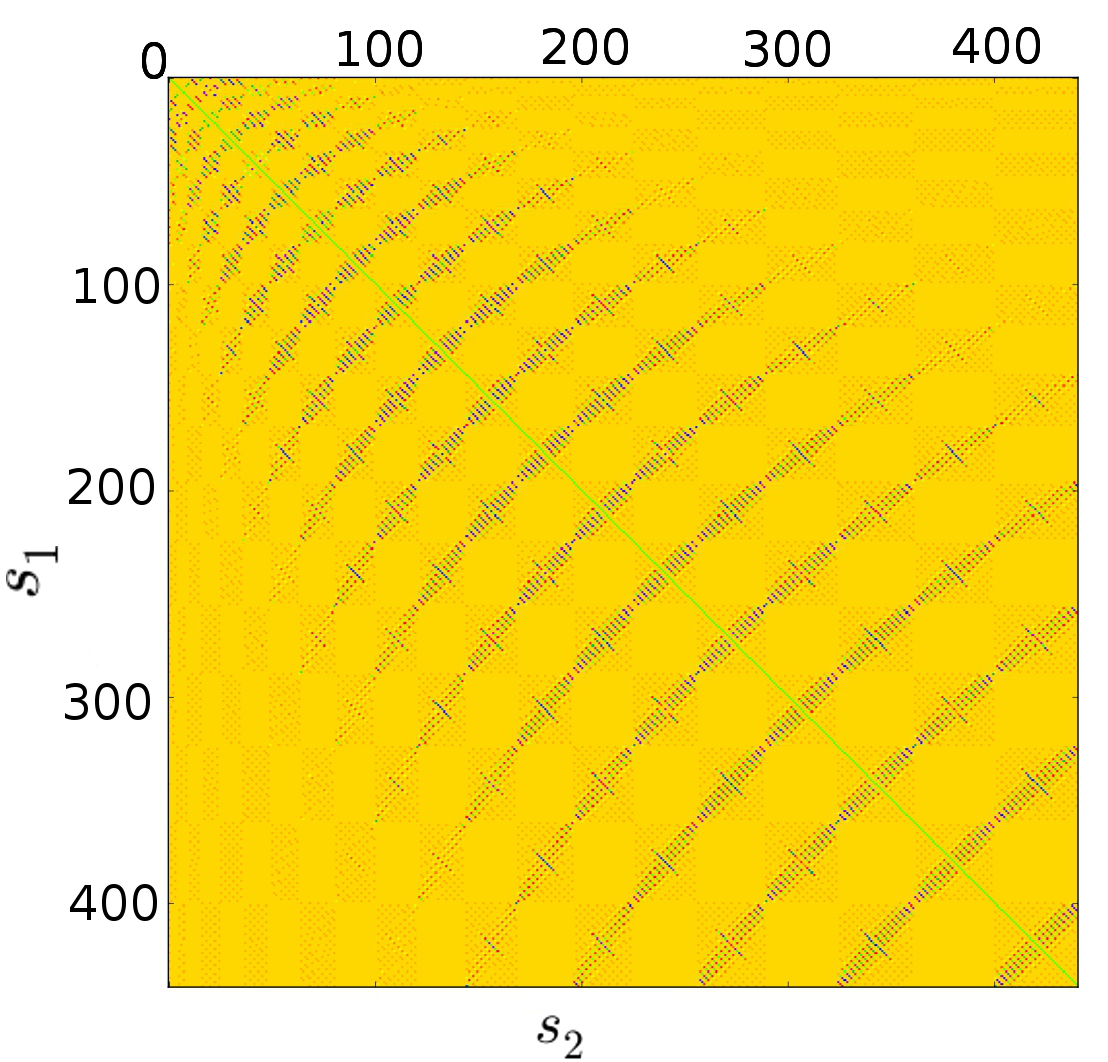}\includegraphics[width=5.8cm]{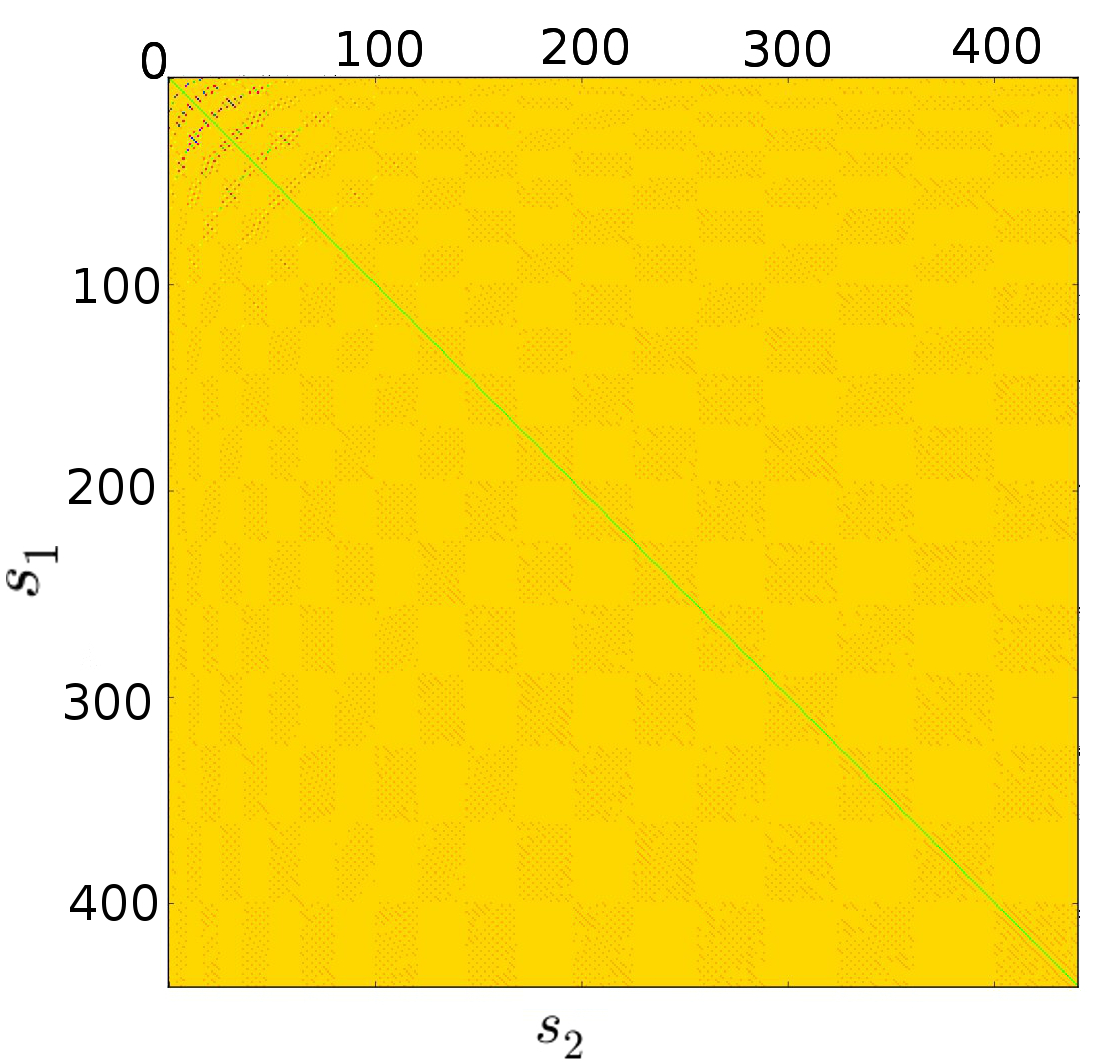}\includegraphics[width=1.2cm]{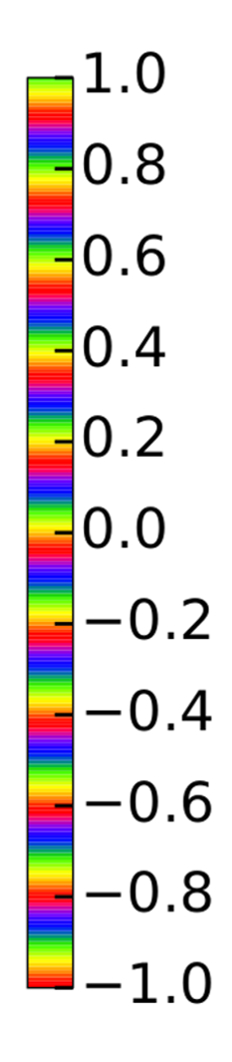}
\caption{Correlation matrices of the $a_{\ell m}$ for a cubic torus of size $L=R_{\mathrm{LSS}}$ (left), $L=D_{\mathrm{LSS}}$ (middle) and $L=1.25D_{\mathrm{LSS}}$  (right).
All assume $\ell_{\rm max}=20$.}
\end{figure*}

Figure~\ref{fig2} presents the correlation matrix for cubic tori of increasing size. We can notice that cubic 3-tori correlation matrices are non-diagonal, because the topology is anisotropic, and block diagonal, due to the properties (\ref{e.34}) and (\ref{torecub}). The non-zero elements are progressively switched off as the size of the 3-torus increases and the correlation matrix looks more like the correlation matrix of a simple Euclidean space, which is a simple diagonal matrix of 1.

\begin{figure*}[htb]\label{fig3}
\centering
\includegraphics[width=5.8cm]{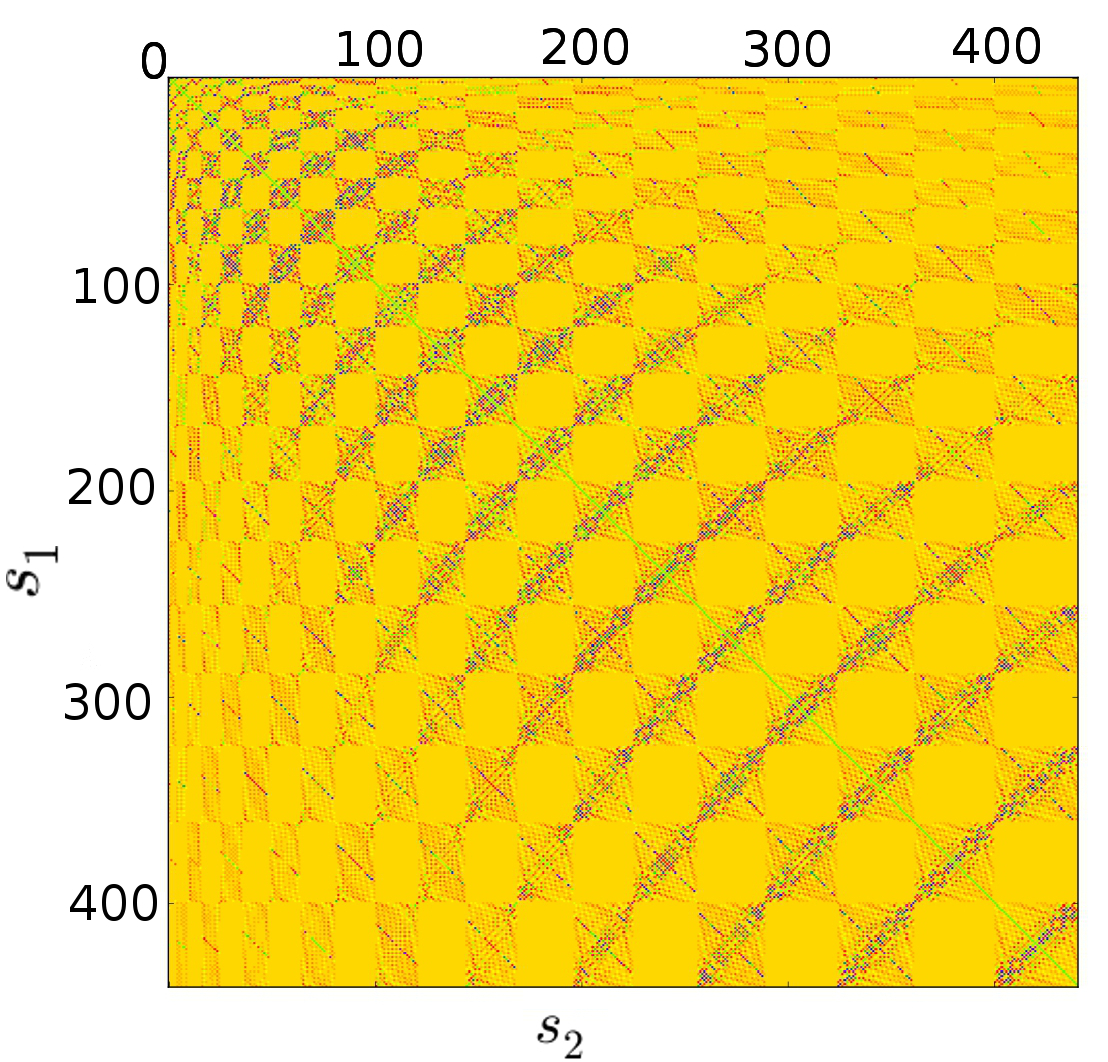}\includegraphics[width=5.8cm]{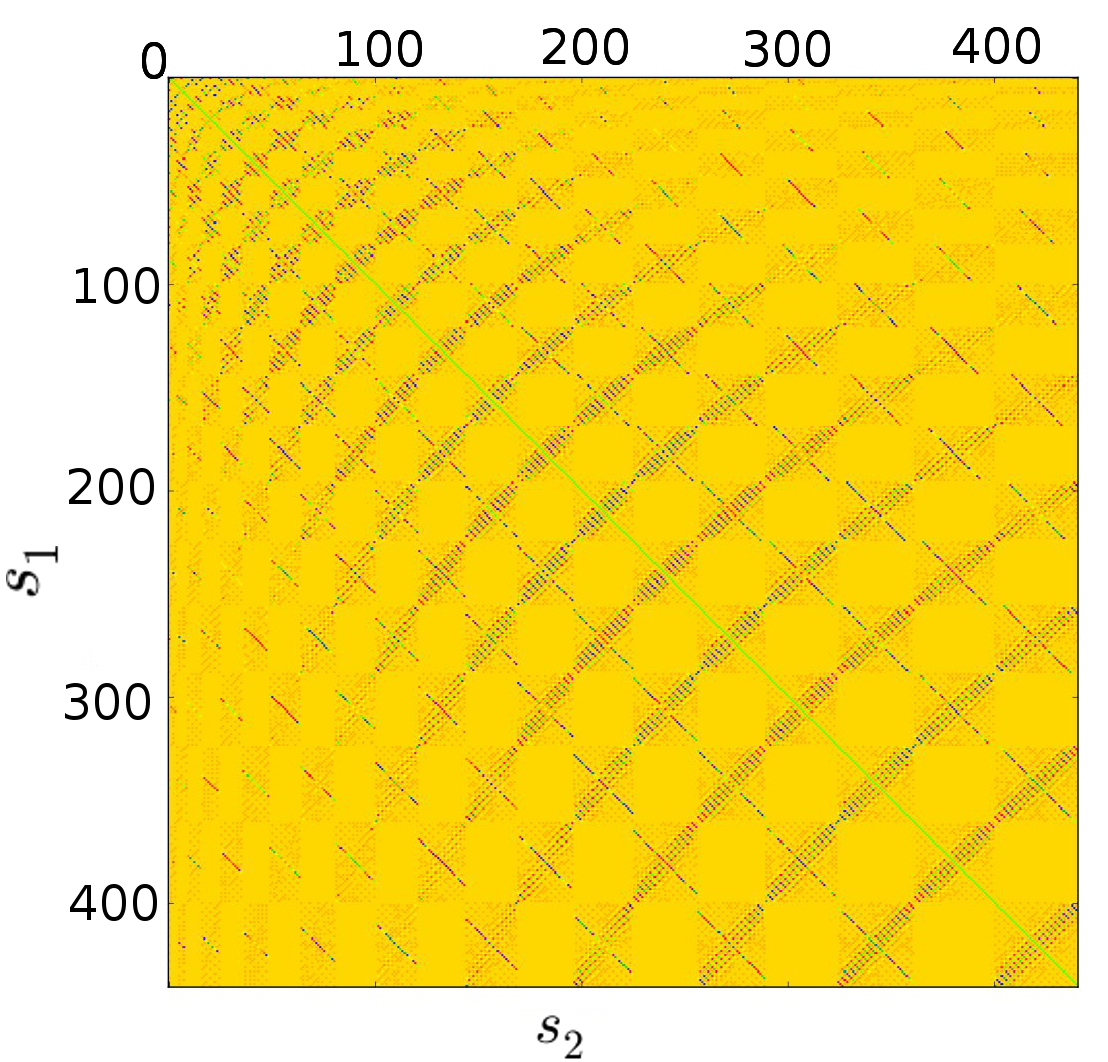}\includegraphics[width=1.2cm]{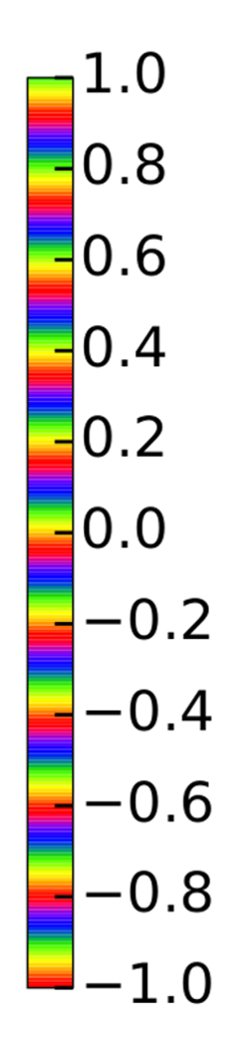}
\caption{Correlation matrices of the $a_{\ell m}$ for a rectangular torus of size $L_x=D_{\mathrm{LSS}}$, $L_y=0.8D_{\mathrm{LSS}}$, $L_z=0.6D_{\mathrm{LSS}}$ (left), and $L_x=L_y=D_{\mathrm{LSS}}$, $L_z=R_{\mathrm{LSS}}$ (right).
All assume $\ell_{\rm max}=20$.}
\end{figure*}

In Fig.~\ref{fig3}, we can see that the block structure of rectangular 3-tori correlation matrices is not exactly the same as in a cubic 3-tori. Besides the modes are differently switched on inside non-zero blocks, especially when $L_x \ne L_y$, $L_x \ne L_z$ and $L_y \ne L_z$, where we can guess rhombic shapes into the blocks. Furthermore rectangular tori have more non-zero modes than cubic ones. This could be a good way to know if we are dealing more likely with a cubic or a rectangular torus, but we should keep in mind that it is just a qualitative approach. Noise and systematic errors will also affect the appearance of the correlation matrix.

\section{Comparing Universe models}\label{section3}

\subsection{Heuristic argument and goal}

\begin{figure*}[htb]\label{fig5}
\centering
\includegraphics[width=6cm]{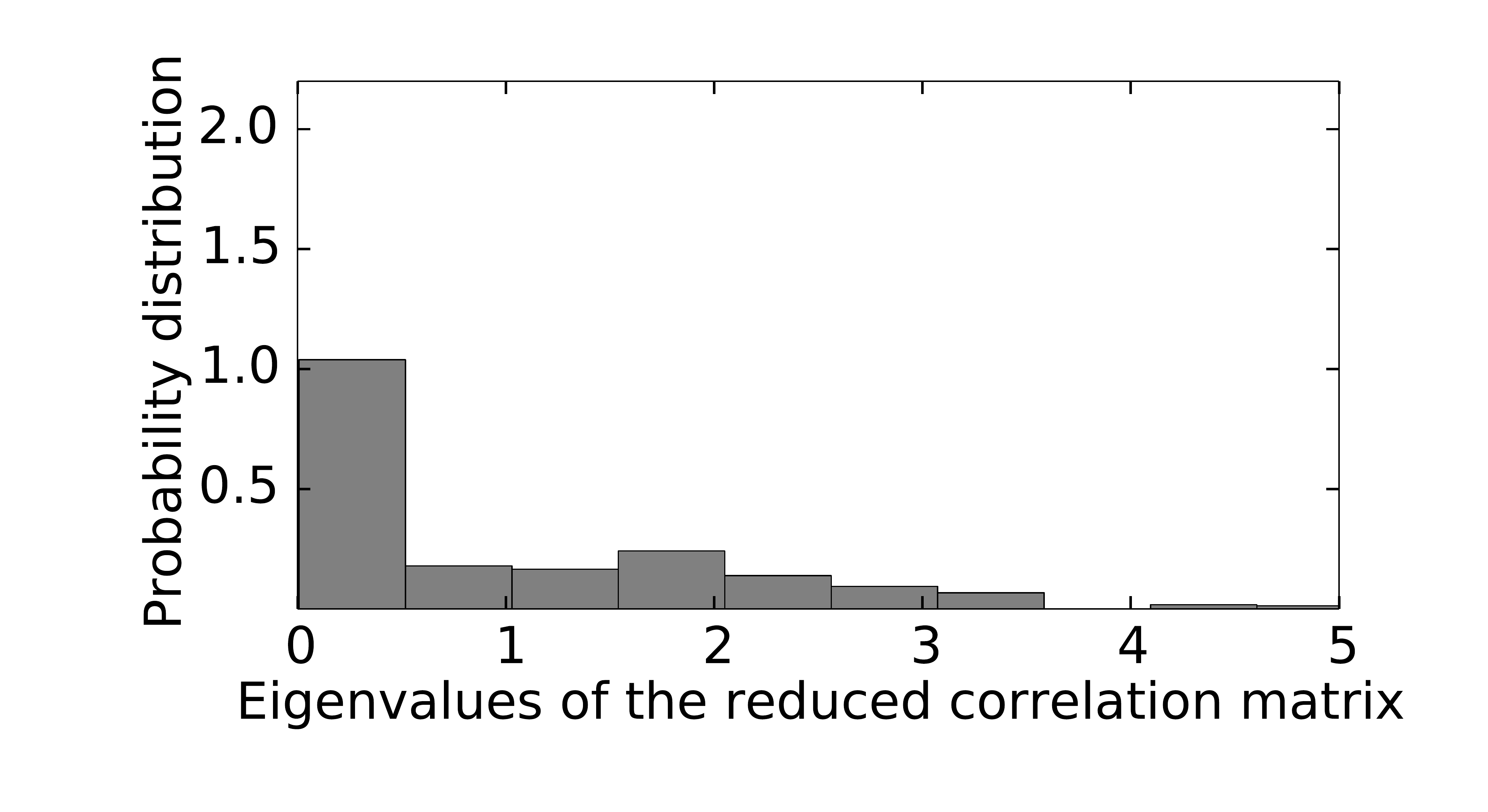}\includegraphics[width=6cm]{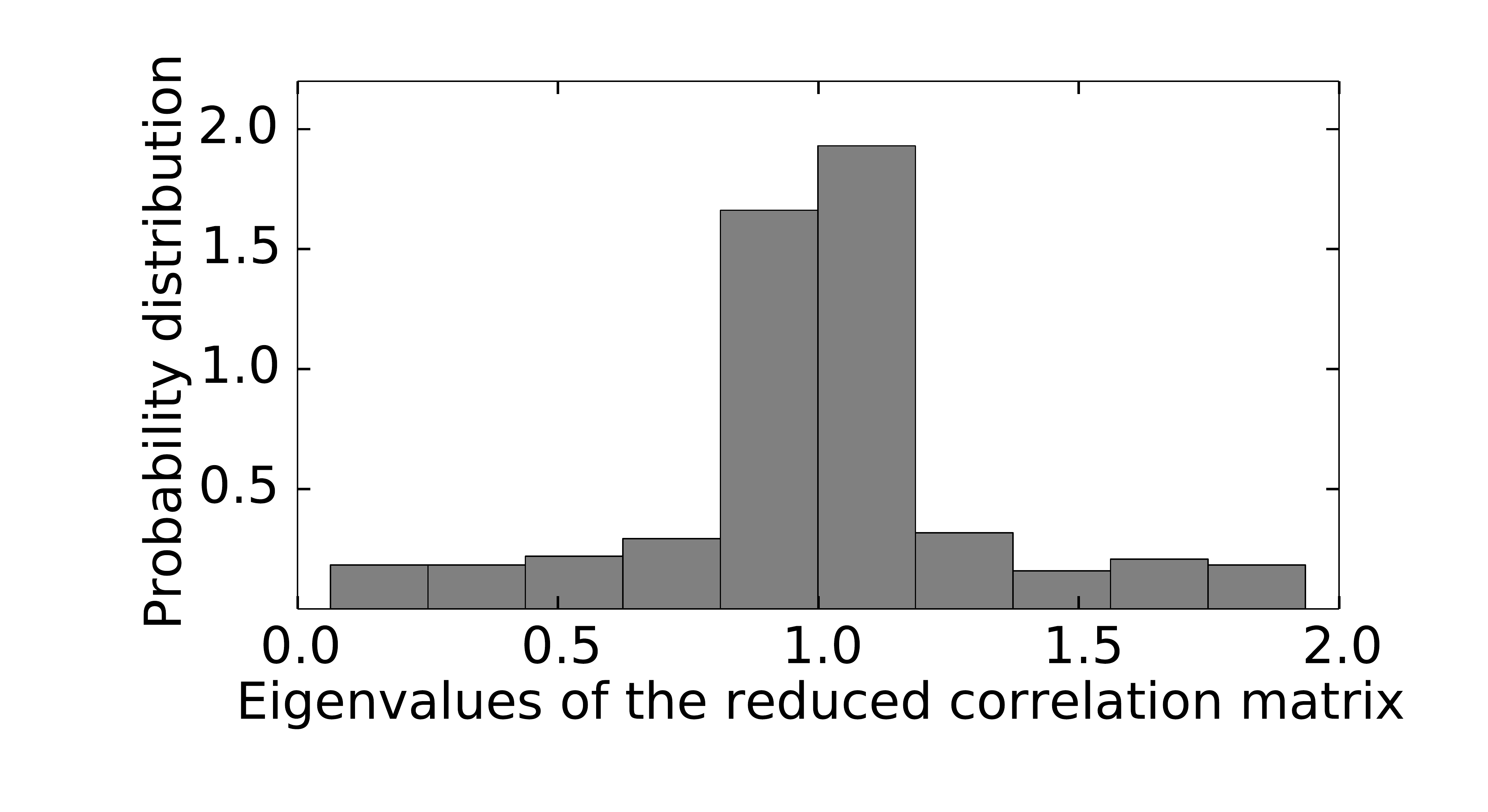}\includegraphics[width=6cm]{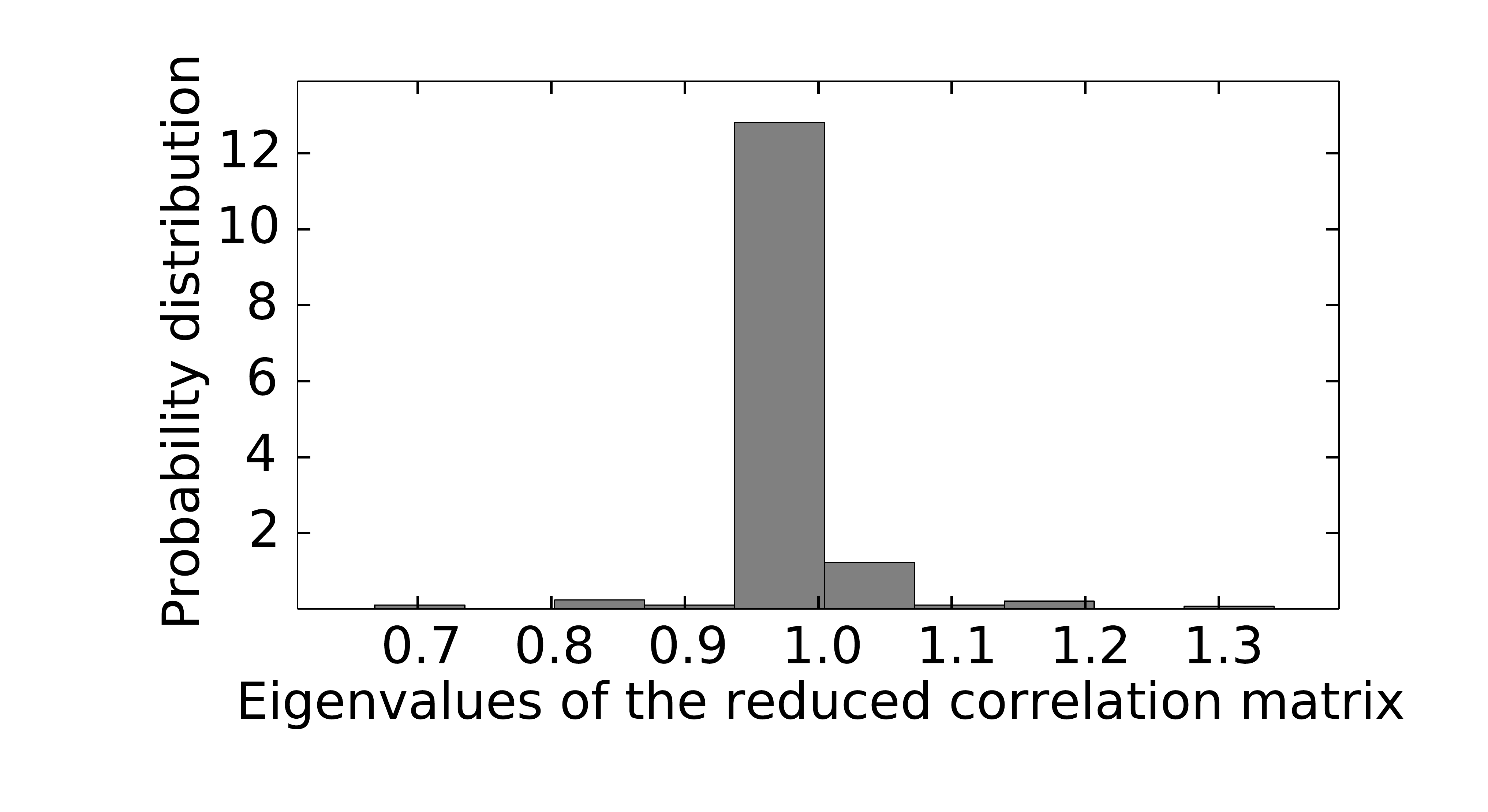}
\caption{Distribution of the eigenvalues of the reduced correlation matrix $A_{ss'}$. The distribution should just be peaked on $1$ for a universe without any topological structure.
(left) $L=R_{\mathrm{LSS}}$ and the mean is $0.984$ and variance $1.07$.
(middle) $L=D_{\mathrm{LSS}}$ and the mean is $0.998$ and variance $0.32$.
(right) $L=1.25 D_{\mathrm{LSS}}$ and the mean is $ 0.9993$ and variance $0.047$. 
All assume $\ell_{\rm max}=\ell_{\rm cut}=20$.}
\end{figure*}

As can be seen by eyes on Fig.~\ref{fig2}, the correlation matrix tends to become more and more diagonal when $L$ increases. Once rescaled by the $C_\ell$ of the isotropic Euclidean space, the normalized correlation matrix $A_{ss'}$ defined by
\begin{equation}
 A_{ss'}=\frac{{C^{(2)}_{ss'}}}{\sqrt{C^{(1)}_\ell C^{(1)}_{\ell'}}},
\end{equation}
where $C^{(1)}_\ell$ is the covariance matrix of the isotropic space and $C^{(2)}_{ss'}$ the covariance matrix of a non-trivial topology, shall converge toward the identity matrix as the size of the non-trivial space increases. This convergence can be visualized heuristically by plotting the distribution of the eigenvalues of the normalized correlation matrix $A_{ss'}$. As can be seen on Fig.~\ref{fig5}, the distribution tends to be more and more peaked around $1$.

\subsection{Kullback-Leibler divergence}

Given the previous discussion, we would like to compare two theories that predict that the coefficients of the expansion of the temperature anisotropies
in spherical harmonics, $a_{\ell m}$, are Gaussian and satisfy
\begin{equation}
\langle a_{\ell m}a_{\ell' m'}^*\rangle_1 = C^{(1)}_{\ell\ell'mm'} = C^{(1)}_\ell \delta_{\ell\ell'}\delta_{mm'}
\end{equation}
for model 1 (isotropic)
and 
\begin{equation}
\langle a_{\ell m}a_{\ell' m'}^*\rangle_2 = C^{(2)}_{\ell\ell'mm'}
\end{equation}
for model 2 (non-trivial topology),
where the ensemble average are taken for each theory respectively.

Such a comparison can be performed in terms of the Kullback-Leibler divergence for two probability distribution functions  $p$ and $q$ defined by
\begin{equation}\label{e.44}
 D_{\mathrm{KL}}(p || q) = \int p(x) \ln\left[\frac{p(x)}{q(x)}\right] \dd x.
\end{equation}
This divergence is the expectation value of $\ln(p/q)$ with the ensemble average related to $p$,
$$
 D_{\mathrm{KL}}(p || q) = \left\langle\ln\left[\frac{p(x)}{q(x)}\right]\right\rangle_p.
$$
Due to the Gibbs inequality, $D_{\mathrm{KL}}$ is always positive. In terms of information theory, $D_{\mathrm{KL}}(p || q)$ quantifies the amount  of information lost when the data $(p)$ is represented by the model $(q)$. Comparing any multi-connected space $(2)$ with the Euclidean trivial space $(1)$ is interesting because the latter has a rotationally invariant covariance matrix. Consequently the  Kullback-Leibler divergence does not depend on the relative orientation of the two spaces and thus quantifies how much information ``separates'' model 2 from model 1. Furthermore the flat Euclidean space is the most probable topology given the previous studies. It is important to see if a deviation from it can be easily detected.
In our case the probability distribution function (PDF) of the $a_{\ell m}$ are given by
\begin{equation}
 \ln P_i(a_{\ell m}) = -\frac{1}{2} {}^ta_{\ell m}\left(C^{(i)}\right)^{-1}_{\ell\ell'mm'}a_{\ell' m'}^* -\frac{1}{2}\ln \det C^{(i)}_{\ell\ell'mm'}
\end{equation}
so that
\begin{eqnarray}
 D_{\mathrm{KL}}(1 || 2) &=& \left\langle\ln\left[\frac{P_1(a_{\ell m})}{P_2(a_{\ell m})}\right]\right\rangle_1.
\end{eqnarray}
\begin{widetext}
Introducing a cut-off $\ell_{\rm cut}$ for the multipole $\ell$, it is explicitely given by
\begin{eqnarray}
 D_{\mathrm{KL}}(1 || 2)                               &=& \frac{1}{2}\ln \left[\frac{\det C^{(2)}_{\ell\ell'mm'}}{\det C^{(1)}_{\ell\ell'mm'}}\right]
                                    + \frac{1}{2}  \left\langle{}^ta_{\ell m}\left(C^{(2)}\right)^{-1}_{\ell\ell'mm'}a_{\ell' m'}^*\right\rangle_1
                                    -  \frac{1}{2}  \left\langle{}^ta_{\ell m}\left(C^{(1)}\right)^{-1}_{\ell\ell'mm'}a_{\ell' m'}^*\right\rangle_1\\
                              &=& \frac{1}{2}\ln \left[\frac{\det C^{(2)}_{\ell\ell'mm'}}{\det C^{(1)}_{\ell\ell'mm'}}\right]  + \frac{1}{2}  \left(C^{(2)}\right)^{-1}_{\ell\ell'mm'}\left(C^{(1)}\right)_{\ell\ell'mm'} -  \frac{1}{2} [ \ell_{\rm cut}(\ell_{\rm cut}+2) - 3] 
\label{formula_kullback}\end{eqnarray}
and we conclude that
\begin{eqnarray}
 2D_{\mathrm{KL}}(1 || 2) = \ln \left[\frac{\det C^{(2)}_{\ell\ell'mm'}}{\det C^{(1)}_{\ell\ell'mm'}}\right]  +  \sum_{\ell=2}^{\ell_{\rm cut}}  C^{(1)}_{\ell}\sum_{m=-\ell}^\ell
 \left(C^{(2)}\right)^{-1}_{\ell\ell mm} - \ell_{\rm cut}(\ell_{\rm cut}+2) + 3.
\end{eqnarray}
As we are interested in cosmological signal only, we start the sum at $\ell=2$ (\emph{i.e.} $s=4$) to get rid of the isotropic component $\ell=0$ and the dipole $\ell=1$. It is easily seen on this  expression that if $C^{(2)}_{\ell\ell'mm'}=C^{(1)}_{\ell\ell'mm'}$ then $D_{\mathrm{KL}}(1 || 2) = 0$. We also notice that in the case of a large cubic torus ($L>D_{\mathrm{LSS}}$), $D_{\mathrm{KL}}(p || q) \approx \frac{1}{2} {\chi}^2$, where ${\chi}^2$ is the usual chi-square distribution.
\end{widetext}

\subsection{Implementation for the topology}

Our previous computation suggests to work with the matrix $A_{ss'}$.  In the case of an isotropic Gaussian distribution it reduces to $\delta_{ss'}$. We now want to estimate how fast does
\begin{equation}
 A_{ss'}\longrightarrow\delta_{ss'}\qquad \hbox{as}\qquad
 L\longrightarrow \infty.
\end{equation}
From the previous expressions, the Kullback divergence is given by
 \begin{eqnarray}
 D_{\mathrm{KL}}(1 || 2) &=& \frac{1}{2}\left[\ln\vert \det A_{s s'}\vert  +  \sum_{s=4}^{\ell_{\rm cut}(\ell_{\rm cut}+2)} 
 \left(A\right)^{-1}_{s s} \right.\nonumber\\
 && \qquad\left.- \ell_{\rm cut}(\ell_{\rm cut}+2)+3\right].
\end{eqnarray}
It takes a simple expression in terms of the eigenvalues $\lambda_i$ of $A_{s s'}$ as
 \begin{eqnarray}
 D_{\mathrm{KL}}(1 || 2) =\frac{1}{2}\sum_i\left[ \ln\vert \lambda_i \vert +  \lambda_i^{-1}-1\right].
\end{eqnarray}
It is obvious on this expression that $D_{\mathrm{KL}}(1 || 2)=0$ when $A_{ss'}$ reduces to the identity. The main interest of this approach is that, unlike the circles in the sky method, one can measure a distance even for spaces with a size larger than $D_{\mathrm{LSS}}$.

\subsection{Detection threshold}

Let us introduce the Bayes factor $B_{12}$ defined as
\begin{equation}
B_{12}=\frac{P_1(d|M_1)}{P_2(d|M_2)}.
\end{equation}
If $B_{12}>1$ (resp. $B_{12}<1$) it represents the increase (resp. decrease) of the credence in favour of model 1 ($M_1$) versus model 2 ($M_2$) given the observed data~\cite{r_trotta}. It gives the factor by which the relatives odds between the two models have changed after taking into account the data. The data are the $a_{\ell m}$ in this experiment.

If we take into account formula (\ref{e.44}), we have
\begin{equation}
 D_{\mathrm{KL}}(1 || 2) = \langle \ln(B_{12}) \rangle_1.
\end{equation}

There is thus a direct link between the Kullback divergence and the Bayes factor. The Jeffrey scale, summarized in Table I, is usually used to interpret the Bayes factor. We can noticed that it is not modified if we consider $\langle\ln(B_{12}) \rangle_1$ instead of $\ln(B_{12})$. As a consequence we obtain the same levels of significance, with a threshold of detectability for $D_{\mathrm{KL}}=1$. This threshold of detectability quantifies the level at which we can distinguish a torus topology from the isotropic model. If $D_{\mathrm{KL}}<1$, the result is inconclusive and the torus topology cannot be distinguished from a Euclidean space. This threshold will be represented in black dotted line in our graphs.

\begin{table}
\begin{tabular}{|l|c|c|c|}
\hline
$|\ln(B_{12})|$ & Odds & Strength of evidence\\
\hline
$<1$ & $<3:1$ & inconclusive  \\
\hline
$1$ & $\approx 3:1$ & weak evidence \\
\hline
$2.5$ & $\approx 12:1$ & moderate evidence \\
\hline
$5.$ & $\approx 150:1$ & strong evidence \\
\hline
\end{tabular}
\label{t.scale}
\caption{Jeffrey scale characterizing the relation between the Bayes factor and the odds.}
\end{table}

\section{How large needs a torus space to be distinguishable from an infinite universe}\label{section4}

\subsection{Ideal experiment: example of a cubic 3-torus}

We have implemented the previous formulae in Python. For a given $\ell_{\rm cut}\leqslant\ell_{\rm max}$, the complexity of the computation of the Kullback-Leibler  divergence scales as $\mathcal{O}({\ell}_{\rm max} {\ell_{\rm cut}}^9 L^3)$.

We present in this section the results for cubic 3-tori, which depend on a single parameter, their size $L$. We let $L$ range from $0.4 D_{\mathrm{LSS}}$ to $1.5 D_{\mathrm{LSS}}$ so that the last tori are larger than the last scattering surface. The correlation matrices $C^{(2)}_{s s'}$ 
include multipoles upto $\ell_{\rm max}=20$ that is upto $s=440$ for all cutting to $\ell_{\rm cut}\leqslant\ell_{\rm max}$.

\begin{figure}[!h]
\includegraphics[width=\columnwidth]{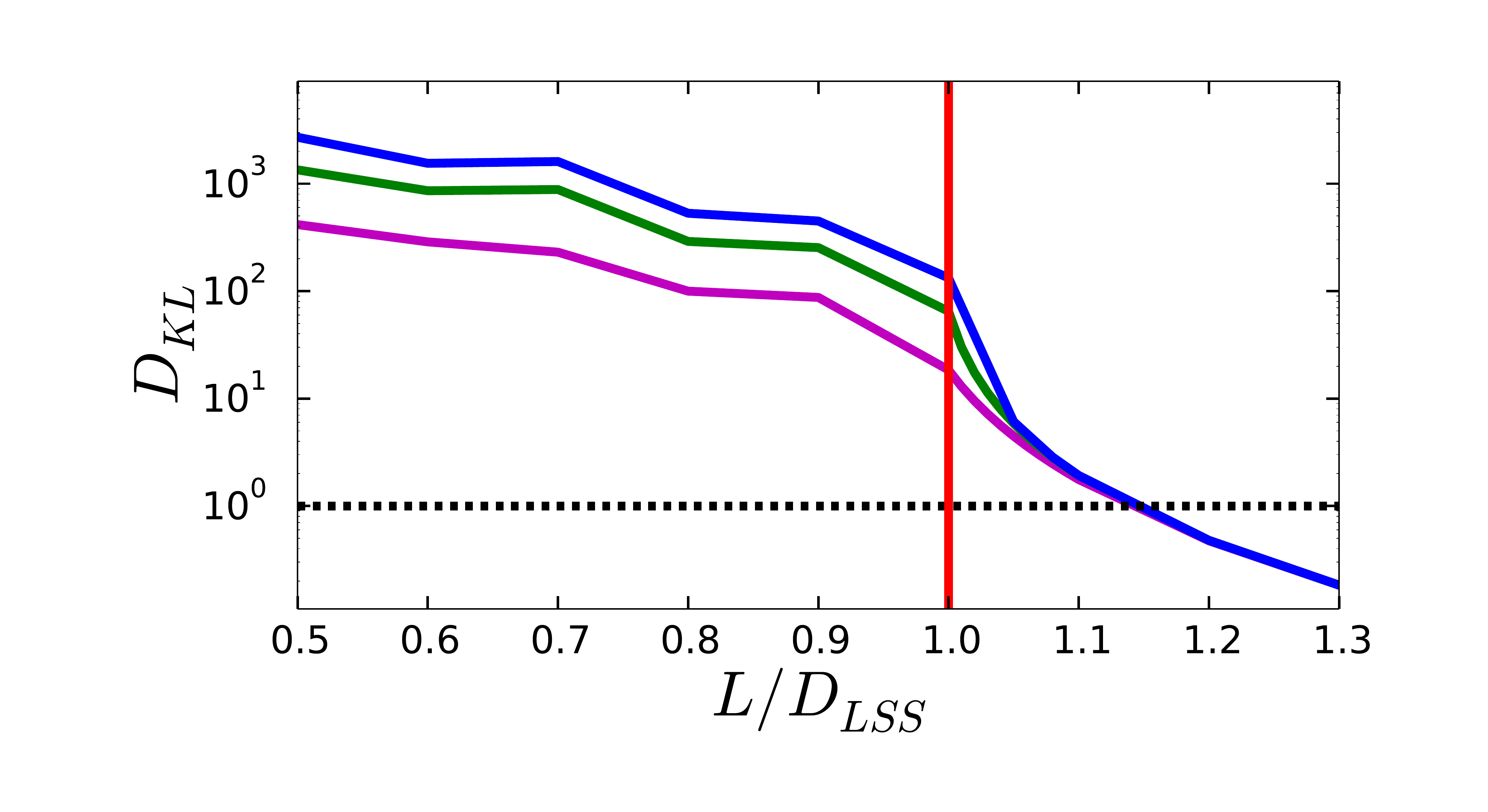}
\caption{Kullback-Leibler divergence at ${\ell}_{\rm max}=30$ for ${\ell}_{\rm cut}=10$ (purple), $20$ (green) and $30$ (blue) as a function of the size of the cubic 3-torus.}
\label{total}
\end{figure}

On the one hand, we clearly see from Fig.~\ref{total}, as expected, that the Kullback divergence $D_{\mathrm{KL}}$ decreases while the size of the 3-torus $L$ increases for a given $\ell_{\rm cut}$. We can distinguish two general behaviors: (1) for $L<D_{\mathrm{LSS}}$, the decrease occurs quite regularly whereas (2) for $L>D_{\mathrm{LSS}}$ there is a dramatic fall-off and a change of slope at $L=D_{\mathrm{LSS}}$. This induces an abrupt difference of $3$ orders of magnitude from the previous regime. 

\begin{figure}[!h]
\includegraphics[width=\columnwidth]{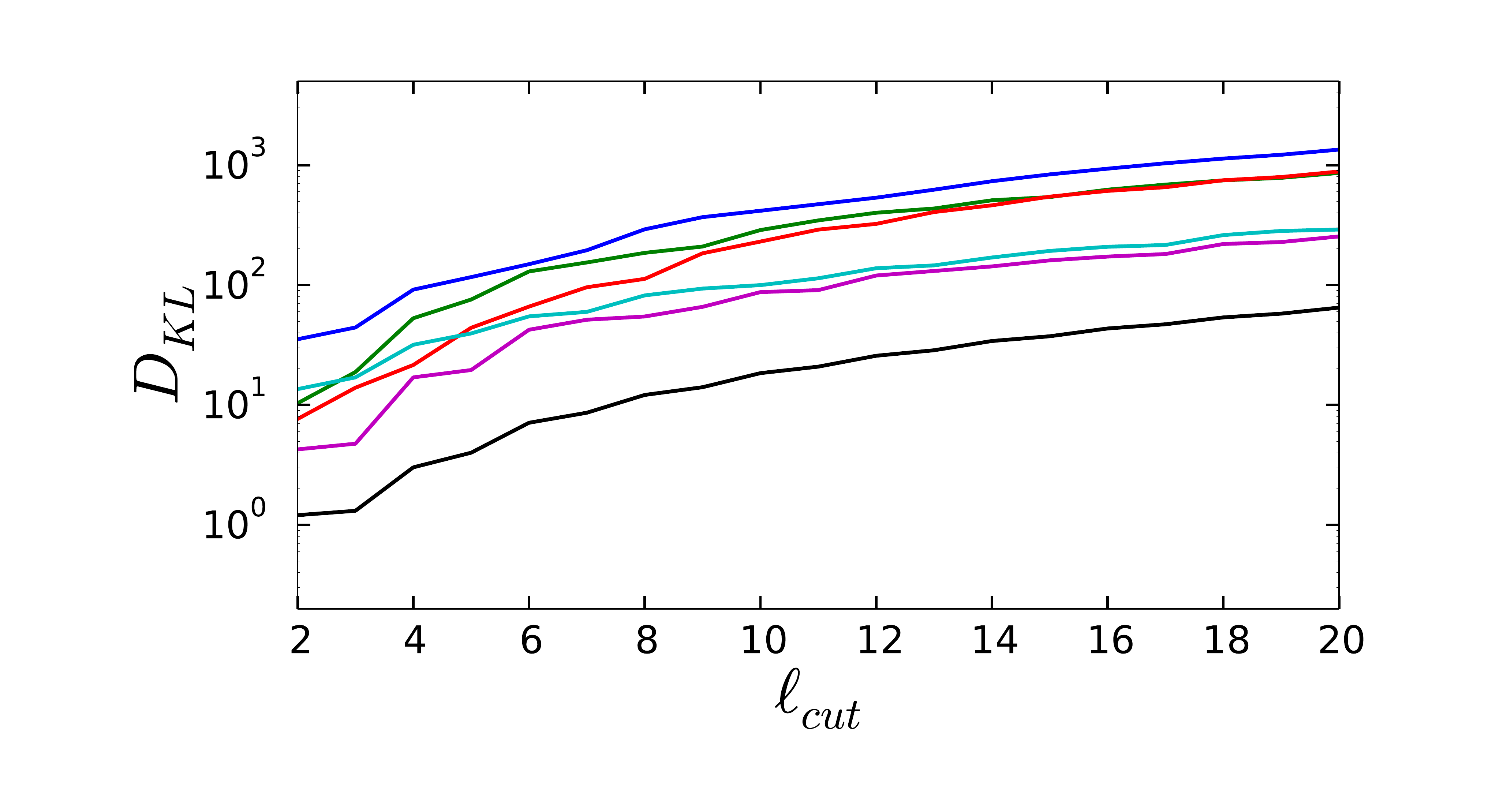}
\caption{Kullback-Leibler divergence for cubic 3-tori of size $L=0.5 D_{\mathrm{LSS}}$ (dark blue), $0.6 D_{\mathrm{LSS}}$ (green), $0.7 D_{\mathrm{LSS}}$ (red), $0.8 D_{\mathrm{LSS}}$ (pale blue), $0.9 D_{\mathrm{LSS}}$ (purple) and $D_{\mathrm{LSS}}$ (black) as a function of $\ell_{\rm cut}$.}
\label{petit_tore}
\end{figure}

\begin{figure}[!h]
\includegraphics[width=\columnwidth]{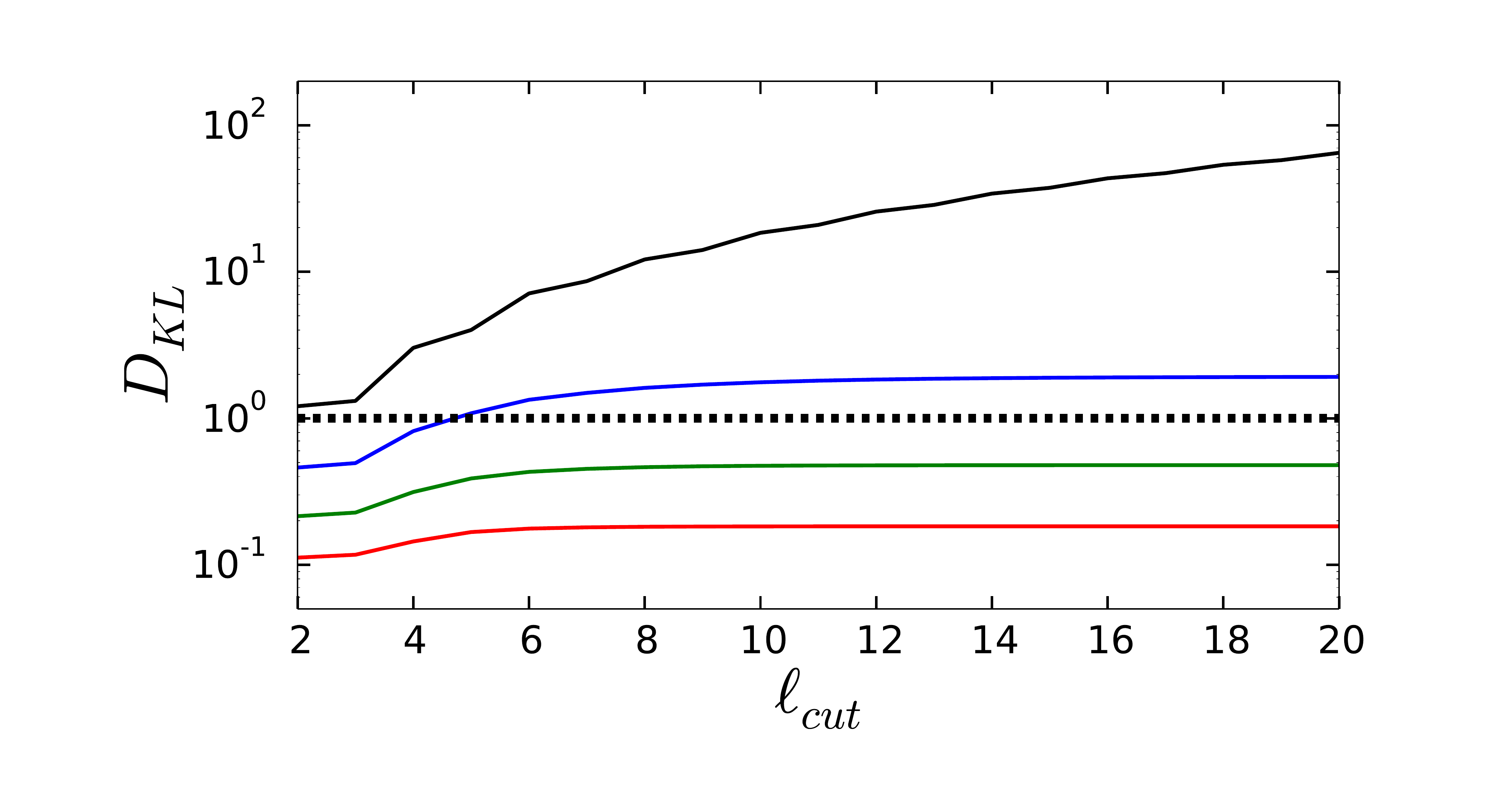}
\caption{Kullback-Leibler divergence for cubic 3-tori of sizes $D_{\mathrm{LSS}}$ (black), $1.1 D_{\mathrm{LSS}}$ (blue), $1.2 D_{\mathrm{LSS}}$ (green) and $1.3 D_{\mathrm{LSS}}$ (red) as a function of $\ell_{\rm cut}$}
\label{grand_tore}
\end{figure}

On the other hand, $D_{\mathrm{KL}}$ seems to increase quadratically with $\ell_{\rm cut}$ for a given fixed $L<D_{\mathrm{LSS}}$ as seen on Fig.~\ref{petit_tore}. When $L>D_{\mathrm{LSS}}$, the curve 
reaches a plateau after a smooth rise for a very small $\ell_{\rm cut}$ as depicted on Fig.~\ref{grand_tore}. For given $\ell_{\rm cut}$ and $L$, there is no sigificant influence of ${\ell}_{\rm max}$ on $D_{\mathrm{KL}}$: the very small variations detected when ${\ell}_{\rm max}$ is increased are only due to the increase of the number of $k$ modes allowed in the $j_\ell$ functions and are negligible.

\begin{figure}[!h]
\includegraphics[width=\columnwidth]{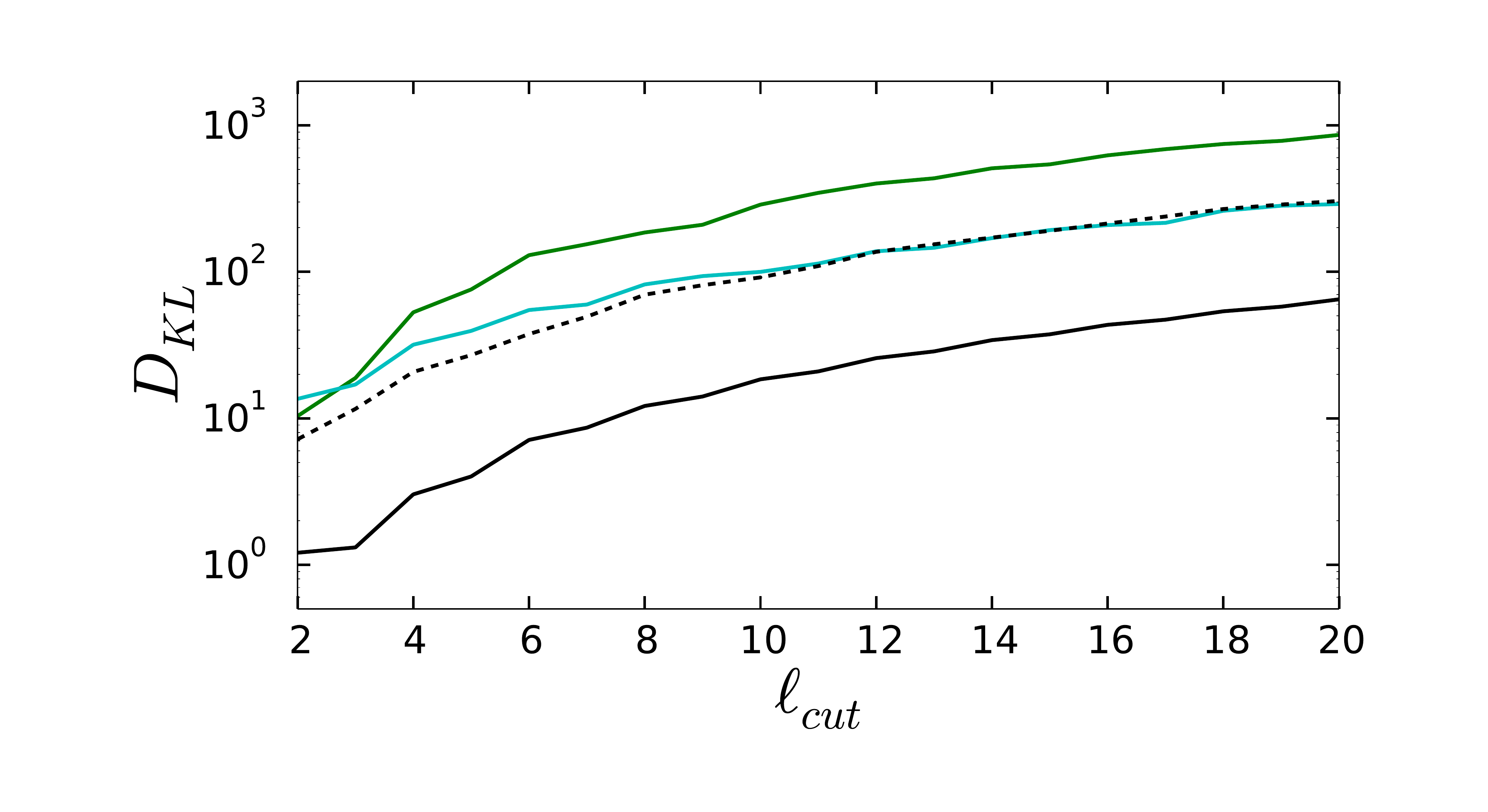}
\caption{Kullback-Leibler divergence for a rectangular 3-tori of sizes for $L_x=0.6 D_{\mathrm{LSS}}$, $L_y=0.8 D_{\mathrm{LSS}}$ and $L_z=D_{\mathrm{LSS}}$ (dotted black) and the associated cubic 3-tori $L=0.6 D_{\mathrm{LSS}}$ (green), $0.8 D_{\mathrm{LSS}}$ (pale blue) and $D_{\mathrm{LSS}}$ (black) as a function of $\ell_{\rm cut}$ for $\ell_{\rm max}=20$.}
\label{rect1}
\end{figure}

\begin{figure}[!h]
\includegraphics[width=\columnwidth]{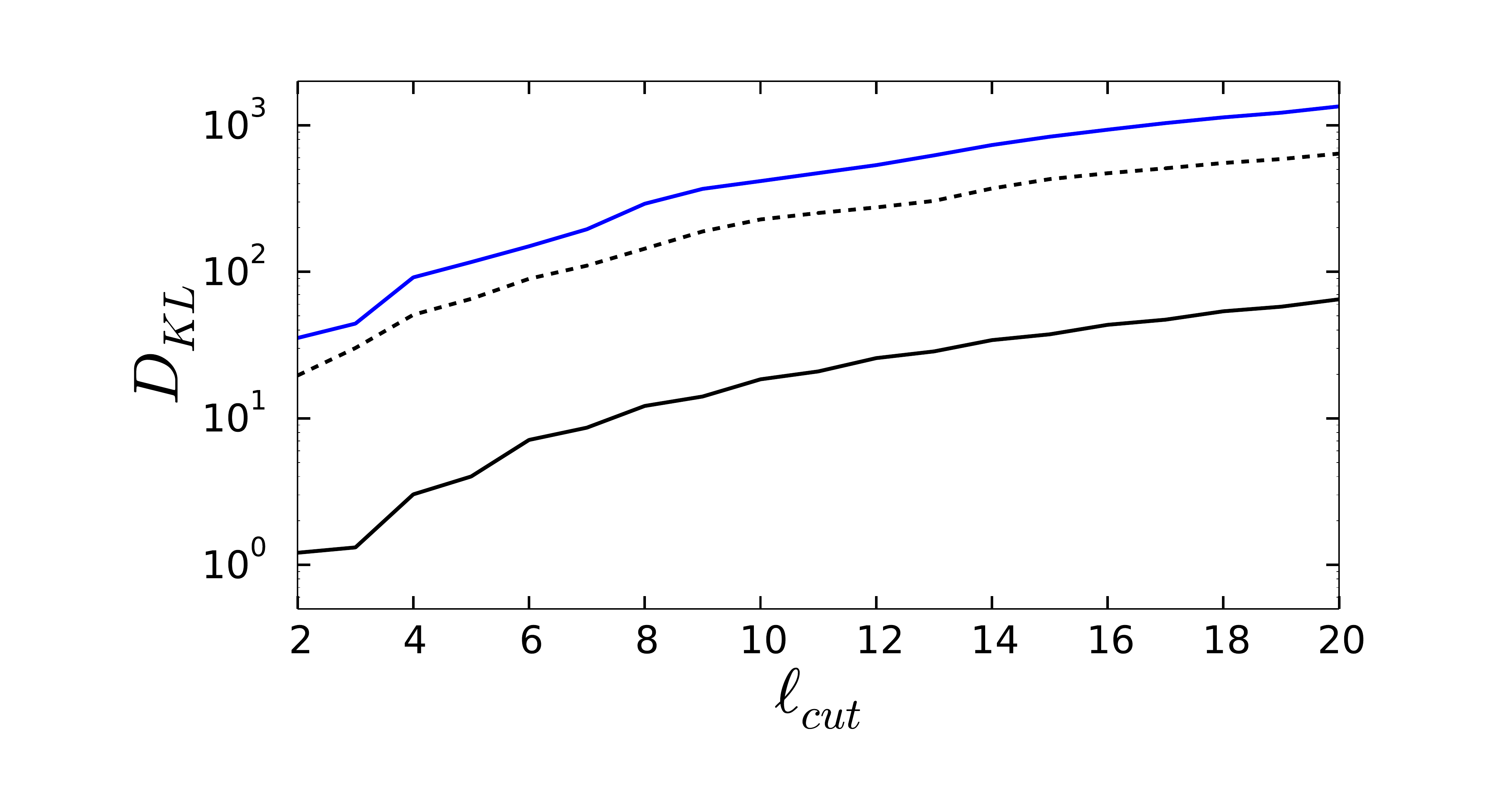}
\caption{Kullback-Leibler divergence for a rectangular 3-tori of sizes for $L_x=L_y=0.5 D_{\mathrm{LSS}}$ and $L_z=D_{\mathrm{LSS}}$ (dotted black) and the associated cubic 3-tori $L=0.5 D_{\mathrm{LSS}}$ (dark blue) and $L = D_{\mathrm{LSS}}$ (black) as a function of $\ell_{\rm cut}$ for $\ell_{\rm max}=20$.}
\label{rect2}
\end{figure}

For a given $\ell_{\rm cut}$, the Kullback-Leibler divergence $D_{\mathrm{KL}}$ decreases as the size of the 3-torus $L$ increases, 
which was intuitively expected as when $L\longrightarrow \infty$, the model gets closer to the isotropic Euclidean 
space, so the difference between the two models becomes thinner.

$D_{\mathrm{KL}}$ also increases with $\ell_{\rm cut}$, \emph{i.e.} when we include more multipoles in the computation: the more $a_{lm}$ taken, the more precise the results are on small scales. 
For large tori, we could have thought that we need to push the computation to high $\ell_{\rm cut}$ in order to obtain a large Kullback-Leibler divergence from Euclidean space, but we notice with Fig.~\ref{total} that increasing $\ell_{\rm cut}$ does not allow us to gain a lot above the threshold of detection. It is thus useless to compute at very high $\ell_{\rm cut}$. This saturation effect implies that for large tori, the dominant part of the information about topology is effectively restricted to large scales. Finally, for smaller tori with $L<D_{\mathrm{LSS}}$, $D_{\mathrm{KL}}$ evolves asymptotically as $\mathcal{O}({\ell_{\rm cut}}^2)$ for big $\ell_{\rm cut}$ roughly following the number of available modes (sum over $\ell_{\rm cut}(\ell_{\rm cut}+2)-3$ terms) but it does not seem that we can write an analytic expression of $D_{\mathrm{KL}}(L,\ell_{\rm cut},{\ell}_{\rm max})$ valid in all regimes. We also perform the same analysis with rectangular tori and we can see from Figs.~\ref{rect1} and ~\ref{rect2} that the results are very similar to the previous results obtained with cubic tori.

The results obtained are qualitatively consistent with those briefly described in the appendix of Ref.~\cite{torus-kunz} although we do not reproduce 
precisely the same results. There are also results very similar to ours in Ref.~\cite{ben2012searching} where the same Kullback-Leiber analysis was performed on a non-classical topology presenting an orbifold point. These references both reproduce the smooth decrease of $D_{\mathrm{LSS}}$ before $L=D_{\mathrm{LSS}}$ and then the sharp decrease for universes bigger than the observable universe.

One major difficulty is the detectability of very big tori spaces. With synthetic ideal data the computation 
can be pushed as far as possible, with computation time limits only. With real data, we are also limited by the resolution of the satellite, which gives an upper bound on ${\ell}_{\rm max}$.
The computation is also limited by a threshold of detection explained in section III.C: if $D_{\mathrm{KL}}<1$, the detection is considered as non-valid. As a consequence, we expect to be able to realistically constrain spaces with tori sizes smaller than $1.15 D_{\mathrm{LSS}}$, as shown in Fig.~\ref{total}. In contrast with the circles-in-the-sky method, which is relevant for $L<D_{\mathrm{LSS}}$ only, our study thus provides a way to investigate spaces with tori larger than $D_{\mathrm{LSS}}$.

One can wonder if the results obtained with only the Sachs-Wolfe contribution of the transfer function are representative of the full transfer function case.
\begin{figure}[!h]
\includegraphics[width=\columnwidth]{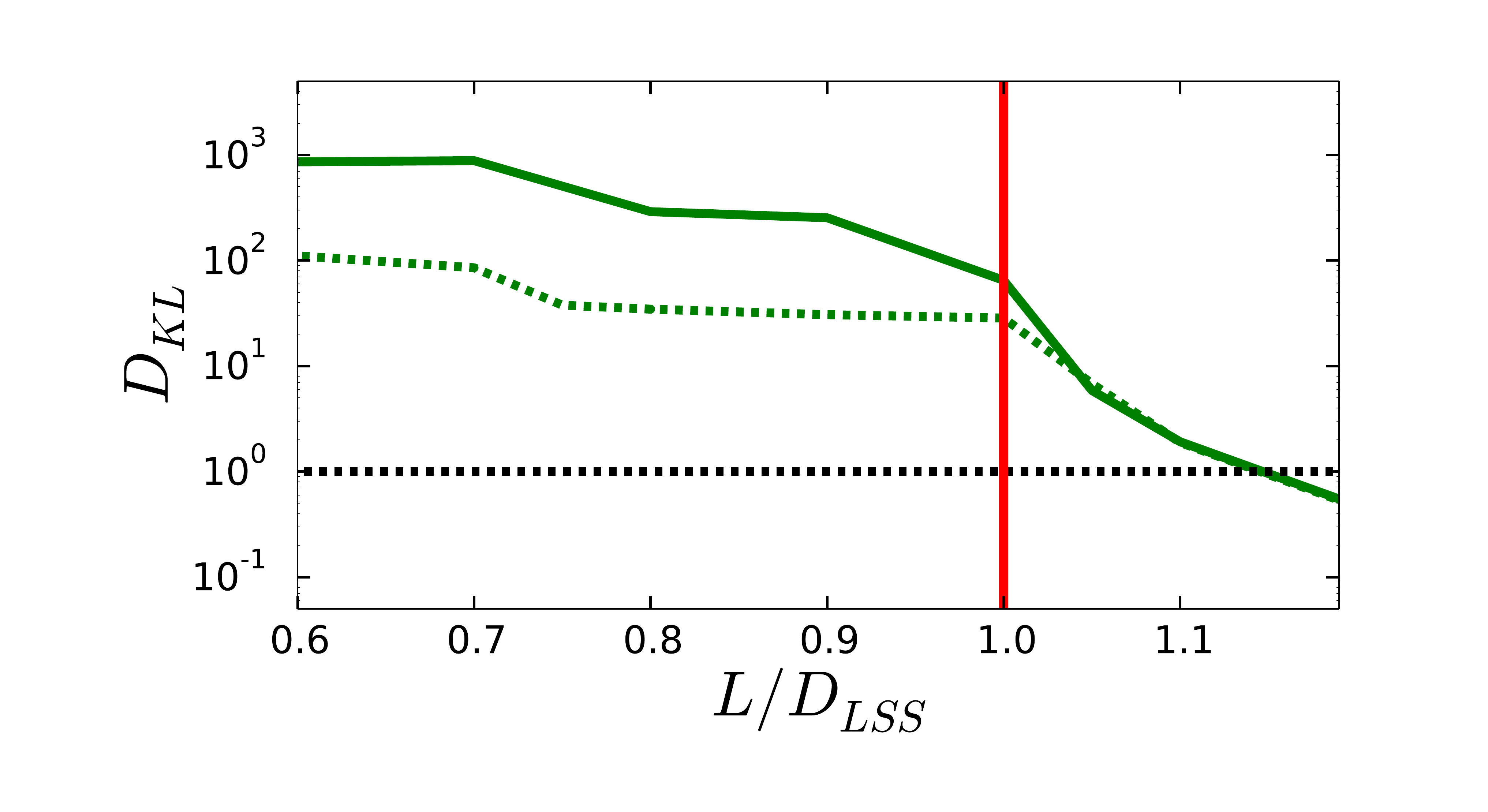}
\caption{Kullback divergence at ${\ell}_{\rm max}=20$ for ${\ell}_{\rm cut}=20$ (green) as a function of the size of the cubic 3-torus with full transfer function in dotted line and only Sachs-Wolfe contribution in plain line.}
\label{dkl_TT_complet}
\end{figure}

As it can be seen on Fig.~\ref{dkl_TT_complet} we get smaller values of $D_{KL}$ for 3-tori smaller than $D_{
\mathrm{LSS}}$ when the full transfer function is used in place of the Sachs-Wolfe approximation. This may be explained by the contribution of the Integrated Sachs-Wolfe (ISW) and Doppler effects at intermediate scales. Indeed, the ISW contribution depends on the photon path from the LSS to the observer, while the Doppler effect depends on the viewing angle of the LSS. Thus, both effects tend to decrease the correlations of matched LSS circles, and overall make it harder to distinguish tori spaces from the Euclidean space. This effect is shown in~\cite{topo-cmb2}: the detection of "circles-in-the-sky" in small tori is excellent with only the Sachs-Wolfe contribution, but if the Doppler effect and the ISW are taken into account, the matched circles are less correlated, and thus more difficult to detect. However, in our study, for 3-tori bigger than $D_{
\mathrm{LSS}}$ the curves are very similar. In the latter regime, the Sachs-Wolfe effect is indeed dominant over the other effects. The largest 3-torus distinguishable from a Euclidean space has size $L_*=1.15D_{
\mathrm{LSS}}$. Once more, there is a similar result in Ref.~\cite{ben2012searching}, where the specific non-classical topology studied is detectable until at least $L=1.1 D_{\mathrm{LSS}}$.

The computation described here is idealized in the sense that it does not take into account 
instrumental noise and foregrounds contaminants, which are major issues of CMB data processing. Their 
impact on $D_{\mathrm{KL}}$ is studied in the following sections.

\subsection{Noise contribution}

So far we have discussed the case of an ideal experiment. In reality, many observational effects, such as e.g. a Galactic cut (to mitigate foreground contamination) or anisotropic instrumental noise, induce non diagonal components to the observed correlation matrix, even in the case of a trivial topology. A purely homogeneous white noise contribution will not induce non diagonal elements, but will make it harder to distinguish between different topologies. The impact of these effects on the detectability of topology therefore needs to be discussed. In the following, we will investigate the impact of a homogeneous white noise component on the Kullback-Leibler divergence. For that purpose, we use noise levels typical of the COBE experiment, as noise levels typical of WMAP or \emph{Planck} are too weak to have an impact on the $D_{\mathrm{KL}}$, which is predominantly sourced by large-scale modes where the CMB anisotropies have largest variance.

Let us redefine the temperature fluctuation as $[\Theta(\Omega)]_{\rm tot} = \Theta(\Omega)+n(\Omega)$ where $n$ is the noise of the satellite. We have, as before,
$$
n_{\ell m}=\int \dd^2\Omega Y_{\ell m}^{*}(\Omega)n(\Omega),
$$
and ${[{a}_{\ell m}]}_{tot}={a}_{\ell m}+n_{\ell m}$, 
$N^{\ell'm'}_{\ell m}=\langle n_{\ell m}n_{\ell'm'}^{*}\rangle$,
${[\CLMLPMP{\ell}{m}{\ell'}{m'}]}_{tot}=\CLMLPMP{\ell}{m}{\ell'}{m'}+N^{\ell'm'}_{\ell m}$ because the temperature fluctuations and the noise are uncorrelated, with $N_\ell \equiv \frac{1}{2\ell + 1} \sum_m {|N_{\ell m}|}^{2}$.

These computations are performed assuming
\begin{equation}
N_\ell=({\Omega}_{\rm pix} {\sigma}_{\rm pix}^2) e^{\ell(\ell+1){\sigma}^2}
\end{equation}
with ${\Omega}_{\rm pix}=\frac{4\pi}{N_{\rm pix}}$ the global solid angle on the map pixels and $N_{\rm pix}$ the number of pixels in the map.

Then we apply our Kullback divergence method on the matrix to get
\begin{equation} 
A_{ss'}=\frac{{[\CLMLPMP{\ell}{m}{\ell'}{m'}]}_{\rm tot}}{\sqrt{(C^{[L\rightarrow \infty]}_\ell +N_{\ell m}^{\ell m}) (C^{[L\rightarrow \infty]}_{\ell'}+N_{\ell'm'}^{\ell' m'})}}
\end{equation}

For a homogeneous white noise contribution, it reduces to
\begin{equation} 
A_{ss'}=\frac{\CLMLPMP{\ell}{m}{\ell'}{m'}+{\delta}_{\ell \ell '}{\delta}_{mm'}N_{\ell }}{\sqrt{(C^{[L\rightarrow \infty]}_\ell +{\delta}_{\ell \ell '}{\delta}_{mm'}N_\ell) (C^{[L\rightarrow \infty]}_{\ell'}+{\delta}_{\ell \ell '}{\delta}_{mm'}N_{\ell'})}}
\end{equation}

\begin{figure}[!h]
\includegraphics[width=\columnwidth]{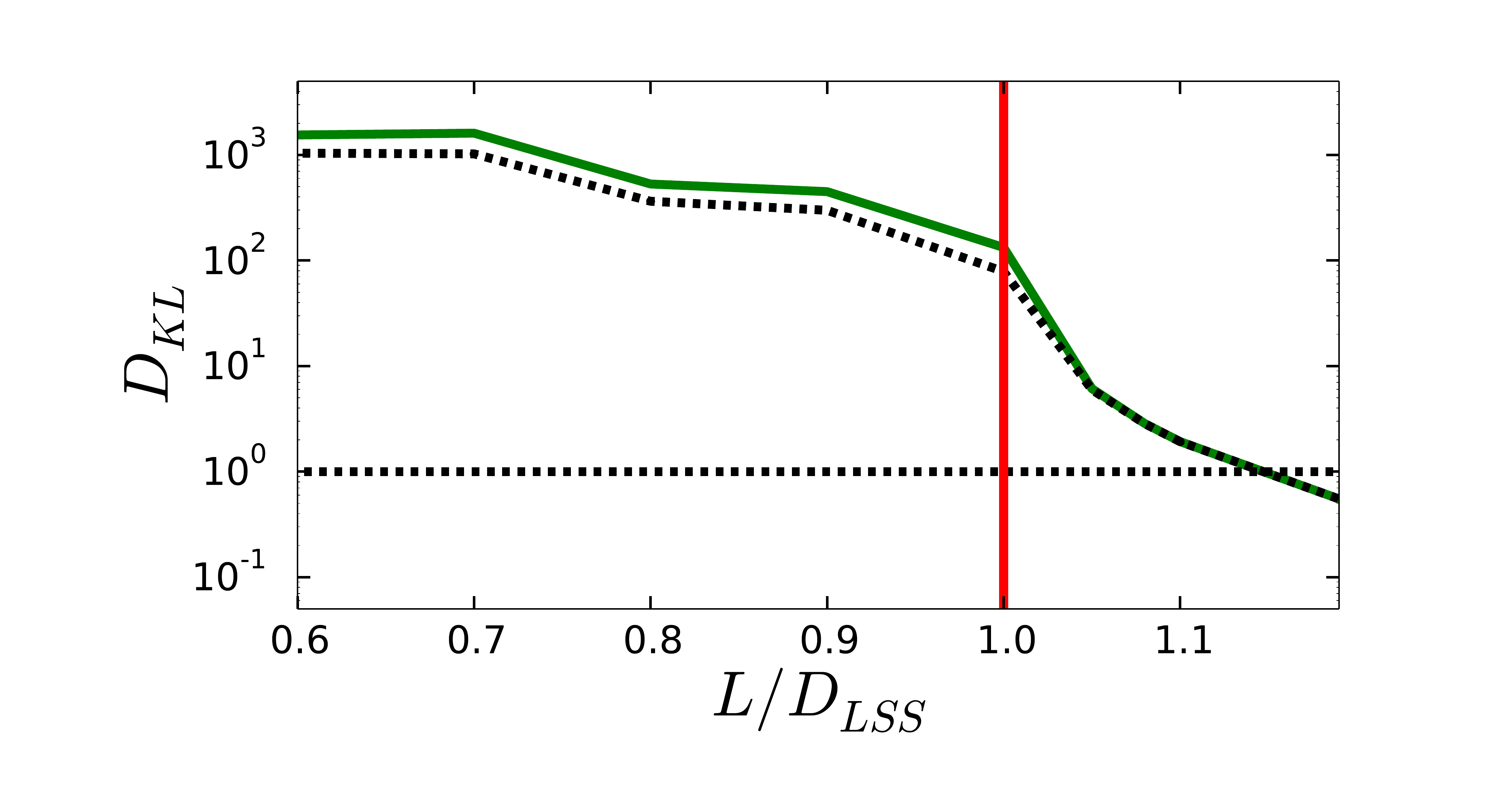}
\caption{Kullback divergence at ${\ell}_{\rm max}=30$ for ${\ell}_{\rm cut}=30$ as a function of the size of the 3-torus without noise (green line) and with COBE noise (black doted line)}
\label{fig:noise}
\end{figure}

We can see in Fig.~\ref{fig:noise} that the effect of the noise on the  Kullback-Leibler divergence remains negligible for large tori, even in the case of a large noise (COBE data). The loss of information is more important for small tori, but remains minor.

\subsection{Galactic mask}

In a realistic situation of CMB data contaminated by Galactic foregrounds, the simplest procedure is to exclude most contaminated part of the sky around the Galactic plane from the analysis. In this section, we investigate the impact of a sky cut on our ability to distinguish between a torus model and an isotropic one.
Masking a part of the sky is equivalent to the application of a projector that is diagonal in real space, hence applying {\it stricto sensu} a rank-deficient matrix on the data. We would therefore naively expect a loss of information proportional to the missing fraction of the sky, and therefore a corresponding decrease of the Kullback divergence $D_{\mathrm{KL}}$.

However, since we are in practice limiting our analysis to low multipoles, we have to investigate the effect of the projector on the vector space spanned by the spherical harmonics up to the maximum multipole considered. As soon as we work in this band-limited setting, it is impossible to get any combination of spherical harmonics with a support strictly contained inside the masked area, and therefore the masking matrix becomes full rank again. We would then expect (again, naively) to get the same results for $D_{\mathrm{KL}}$ as in the full-sky ideal case, since no information is lost in principle.
This is not the case however, as masking adds an additional coupling between the multipoles, and effectively transfers power to higher multipoles. Therefore, at a fixed maximum multipole, we expect a loss of information due to this power transfer to scales that are not considered in the analysis.

\begin{figure}[!h]
\includegraphics[width=6cm,height=5cm]{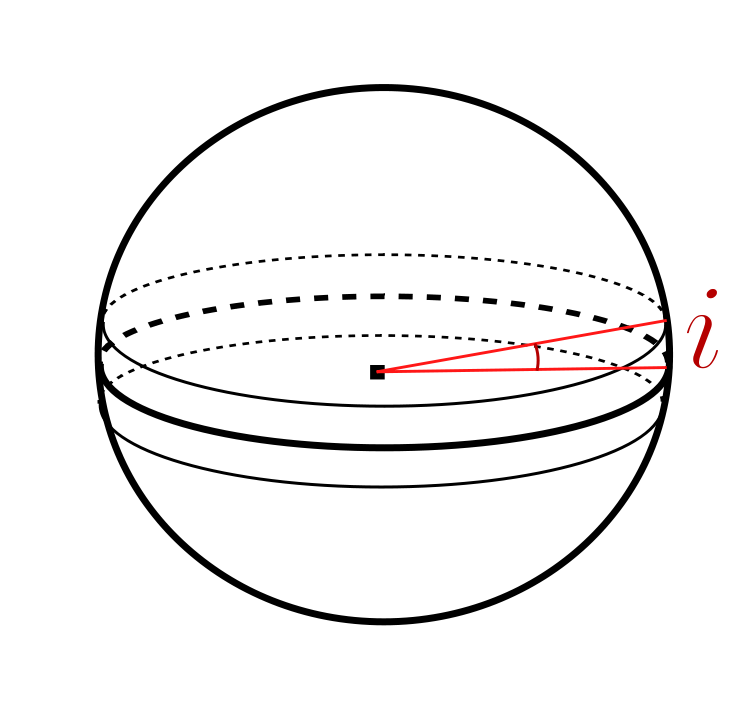}
\includegraphics[width=6cm,height=5cm]{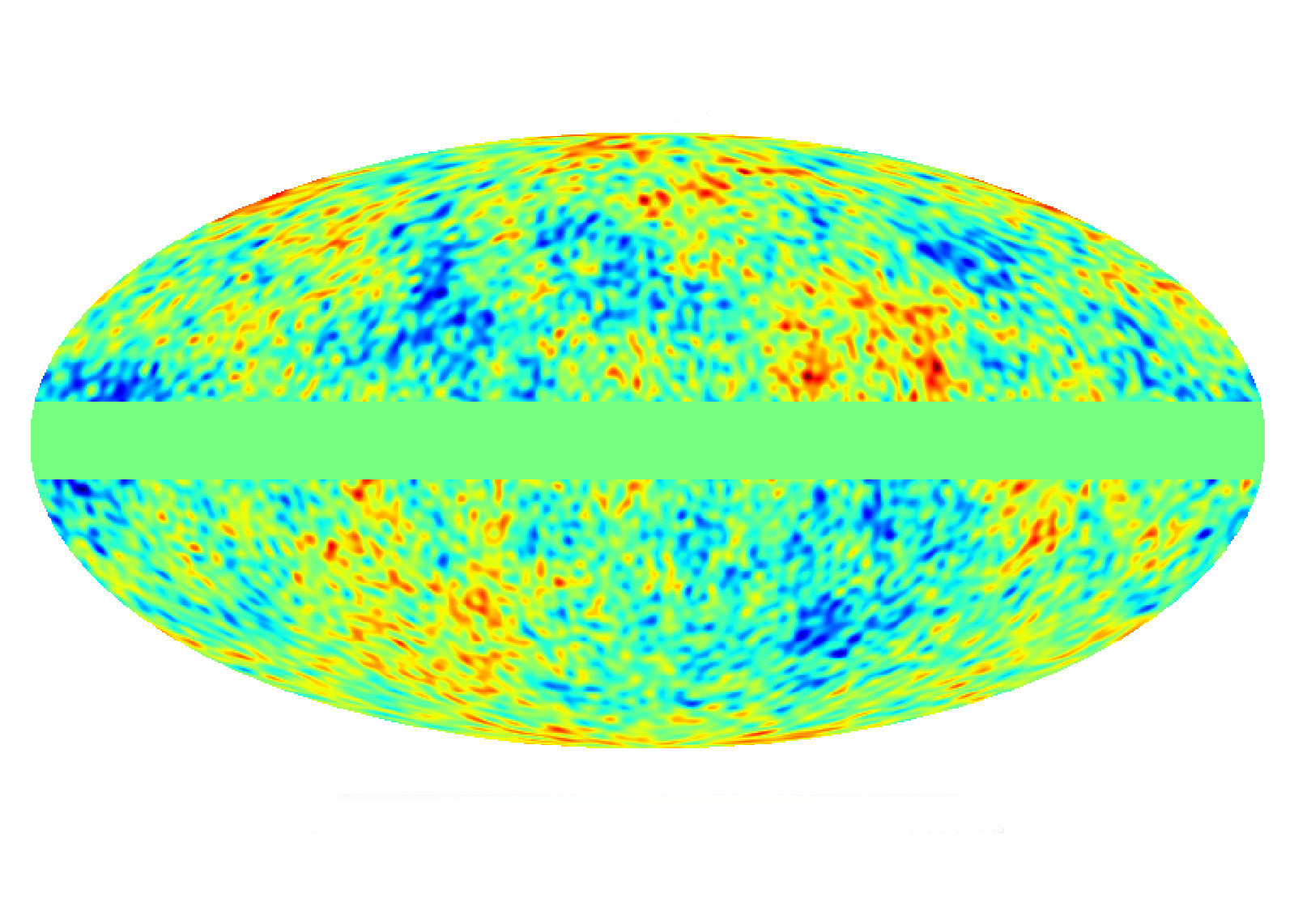}
\caption{The form of the mask for the sky cut is taken to be very simple. It reduces to an azimuthal strip of colatitude $i$ (top), so that the map used foe the analysis is given by the bottom figure.}
\label{coupure}
\end{figure}

For simplicity, we consider here an azimuthally symmetric Galactic mask centered on the equatorial plane (see Fig.~\ref{coupure}). 
Our former covariance matrices are modified as follow
$C^{(1)} \rightarrow KC^{(1)}K^{T}$ and $C^{(2)} \rightarrow KC^{(2)}K^{T}$, where $K$ is the matrix related to the mask

\begin{equation}
K_{\ell m}^{\ell'm'}=\int d \Omega Y^{*}_{\ell m}(\Omega)M(\Omega)Y_{\ell'm'}(\Omega)
\end{equation}
if we take $M(\Omega)$ the function describing the effect of the mask on the sphere

\begin{equation}\label{eq:mask}
M(\theta,\phi)=\left\{ \begin{array}{ll} 
                1 & \mbox{ if }  |\theta-\pi/2|<i \\
                0&  \mbox{ else} \\
          \end{array}\right.
\end{equation}

In the particular case of an azimuthal cut, the matrix $K$ does not couple azimuthal modes Ref.~\cite{master-paper}

\begin{equation}
K_{\ell m}^{\ell'm'}=K_{\ell m}^{\ell'm} {\delta}_{mm'}
\end{equation}

\begin{figure}[!h]
\includegraphics[width=\columnwidth]{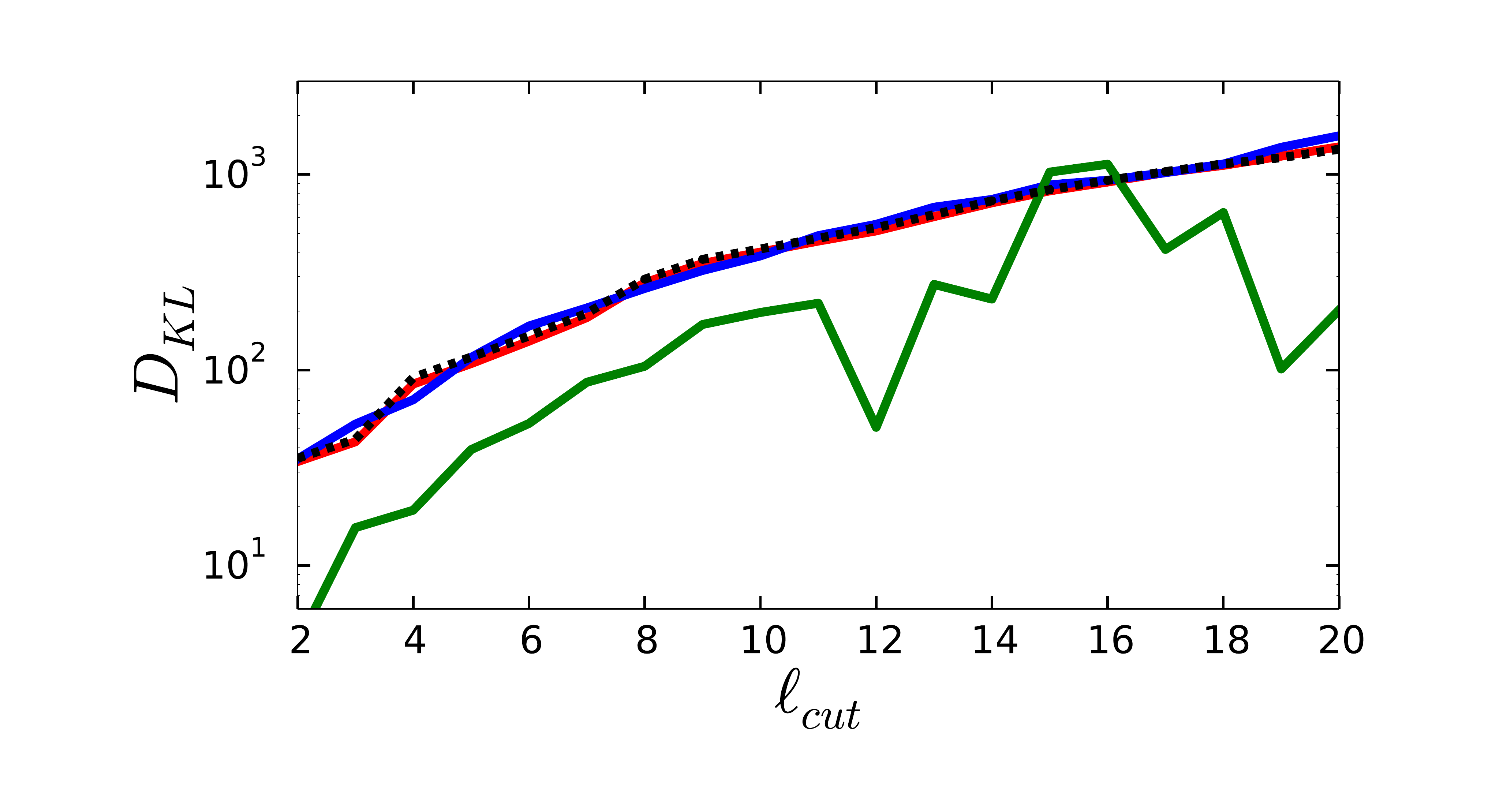}
\caption{Impact of masking out data on the Kullback-Leibler divergence $D_{\mathrm{KL}}$ between a cubic torus model of size $0.5 D_{\mathrm{LSS}}$ (black), as a function of the analysis bandwidth $\ell_{\rm cut}$, and for different mask sizes, respectively for $i=60^{\circ}$ (green), $18^{\circ}$ (blue), and $3.6^{\circ}$ (red), as defined in equation~\ref{eq:mask}.The reference case with } 
\label{coupure_graphe}
\end{figure}

We depict on Figure~\ref{coupure_graphe} the net impact of masking out data on $D_{\mathrm{KL}}$, for different $\ell_{\rm cut}$ and different sizes of masks. We observe, as expected, a noticeable decrease of $D_{\mathrm{KL}}$ for large masks, but a negligible effect for smaller masks. This is compatible with the results of Ref.~\cite{efsta} who found that for sufficiently small masks and low multipoles, there is effectively no loss of information compared to the full-sky case.

\section{Polarization}\label{section5}

Up to now we have only considered CMB temperature anisotropies. Another source of information is the polarization of the CMB, even if only $10\%$ of the anisotropies are expected to be polarized at maximum~\cite{polarization}. The CMB is almost unpolarized before decoupling but Thompson scattering tends to linearly polarize the radiation in the direction normal to the surface of diffusion. This effect only occurs if the radiation is anisotropic before scattering , with a quadrupolar anisotropy~\cite{pubook,polarization}. The goal of this section is to investigate if the additional information brought by polarization do improve or not the detection of non-trivial topologies.
After a briefly describing polarization formalism on the sphere, we show the effect of a torus topology on the polarized power spectra. We then generalize the computation of the Kullback-Leibler divergence to include polarization information, and show how the detectability of tori of size $L<D_{\mathrm{LSS}}$ is enhanced in the case of an ideal experiment.

 \subsection{Stokes parameters}

Let us consider an electromagnetic wave

\begin{equation}
\vec{E}=\vec{E_0}e^{i(\omega t -kz)}
\end{equation}
 with
\begin{equation}
 \vec{E_0}=E_x \vec{e_x}+E_y \vec{e_y},\  E_x=A_1e^{-i{\Theta}_1}\  \text{and}\  E_y=A_2e^{-i{\Theta}_2}.
\end{equation}

We define the four Stockes parameters as Refs.~\cite{pubook,polarization}

\begin{equation}
I=\langle A_1^2\rangle+\langle A_2^2\rangle,
\end{equation}
 which represents the total intensity of the wave,

\begin{equation}
Q=\langle A_1^2\rangle-\langle A_2^2\rangle,
\end{equation}
which measures the excess of linear polarization in the $x$ direction compared to the $y$ direction,

\begin{equation}
U=2\langle A_1A_2\mathrm{cos}(\Theta_1-\Theta_2)\rangle,
\end{equation}
which is determined via $I^2=Q^2+U^2$ and is a caracterization of linar polarization too, and

\begin{equation}
V=2\langle A_1A_2\mathrm{sin}(\Theta_1-\Theta_2)\rangle
\end{equation}
which gives the difference between the positive and the negative helicities and is thus related to circular polarization. If there is no initial circular polarization in the radiation Thompson scattering will not generate any. That is why we do not consider $V$. Furthermore $I$ is completely deduced from $U$ and $Q$. As a consequence it is sufficient to study CMB polarization to consider only $Q$ and $U$ which characterize entirely the polarization field . The interesting quantity to study is $[Q\pm iU]$~\cite{pubook,polarization}.
We can thus make an analogy with formula (17) by projecting these functions on the adequate spherical harmonics basis as

\begin{equation}
[Q\pm iU](\theta, \varphi)  = \sum_{\ell = 0}^\infty \sum_{m = - \ell}^\ell
a_{lm}^{(\pm 2)} Y_{\ell m}^{(\pm 2)} (\theta, \varphi)
\end{equation}

$Q$ and $U$ are real numbers, so ${a_{lm}^{(-2)}}^{*}=a_{l-m}^{(2)}$. 

\subsection{$E$-modes and $B$-modes}

Let us introduce

\begin{equation}\label{dttalm1}
E_{lm}=-\frac{1}{2}\left(a_{lm}^{(2)}+a_{lm}^{(-2)}\right)
\end{equation}

and

\begin{equation}\label{dttalm2}
B_{lm}=\frac{i}{2}\left(a_{lm}^{(2)}-a_{lm}^{(-2)}\right)
\end{equation}

We thus have two non-local parameters in real space $E$ and $B$ describing totally the polarization field

\begin{equation}\label{dttalm3}
E(\theta, \varphi)=\sum_{\ell = 0}^\infty \sum_{m = - \ell}^\ell
E_{lm} Y_{\ell m}(\theta, \varphi)
\end{equation}

and

\begin{equation}\label{dttalm4}
B(\theta, \varphi)=\sum_{\ell = 0}^\infty \sum_{m = - \ell}^\ell
B_{lm} Y_{\ell m}(\theta, \varphi).
\end{equation}

There has been, to the present day, no detection of $B$-modes. Theoretically $E$-modes are generated by scalar, vector and tensor perturbations whereas $B$-modes are generated by vector and tensor perturbations. 

In Ref.~\cite{polarization-wmap}, the limit of the tensor-to-scalar ration $r$ evaluated with the WMAP data combined with 
BAO and SN is found to be $r<0.22$ with a $95\%$ confidence level. More recently the \emph{Planck} collaboration~\cite{Planck_inflation} has established an upper boundary $r<0.11$ with a $95\%$ confidence level. As a consequence, even if $B$-modes are supposed to take part in the CMB fluctuations, their contribution 
is negligible compared to $E$-modes. That is why in the remaining part of the study, we will only consider $E$-modes as in Ref.~\cite{polarization-topology}.

 \subsection{Power spectra}

The full transfer functions for temperature fluctuations and polarization have been obtained from the CAMB software Ref.\cite{CAMB}.

\begin{figure}[!h]
\includegraphics[width=\columnwidth]{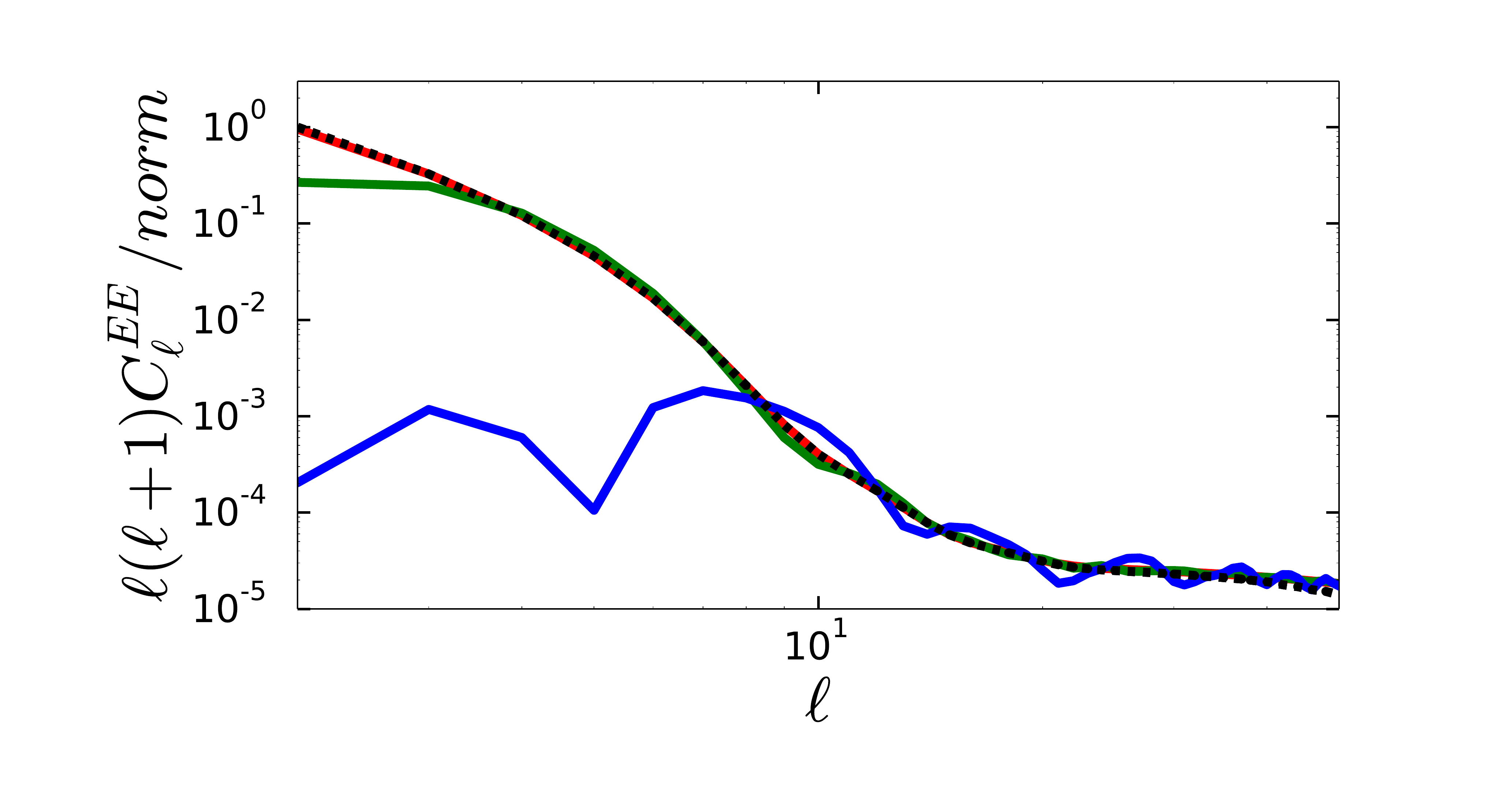}
\caption{$EE$ angular power spectrum for 3-tori of size $L=0.25D_{
\mathrm{LSS}}$ (blue), $L=0.5D_{
\mathrm{LSS}}$ (green), $L=D_{
\mathrm{LSS}}$ (red) and for the Euclidean space (dotted black line). The computation was done by taking into account the full transfer functions from CAMB. The norm is taken equal to $6C_{2}^{EE,\mathrm{iso}}$.}
\label{cl_EE}
\end{figure}

\begin{figure}[!h]
\includegraphics[width=\columnwidth]{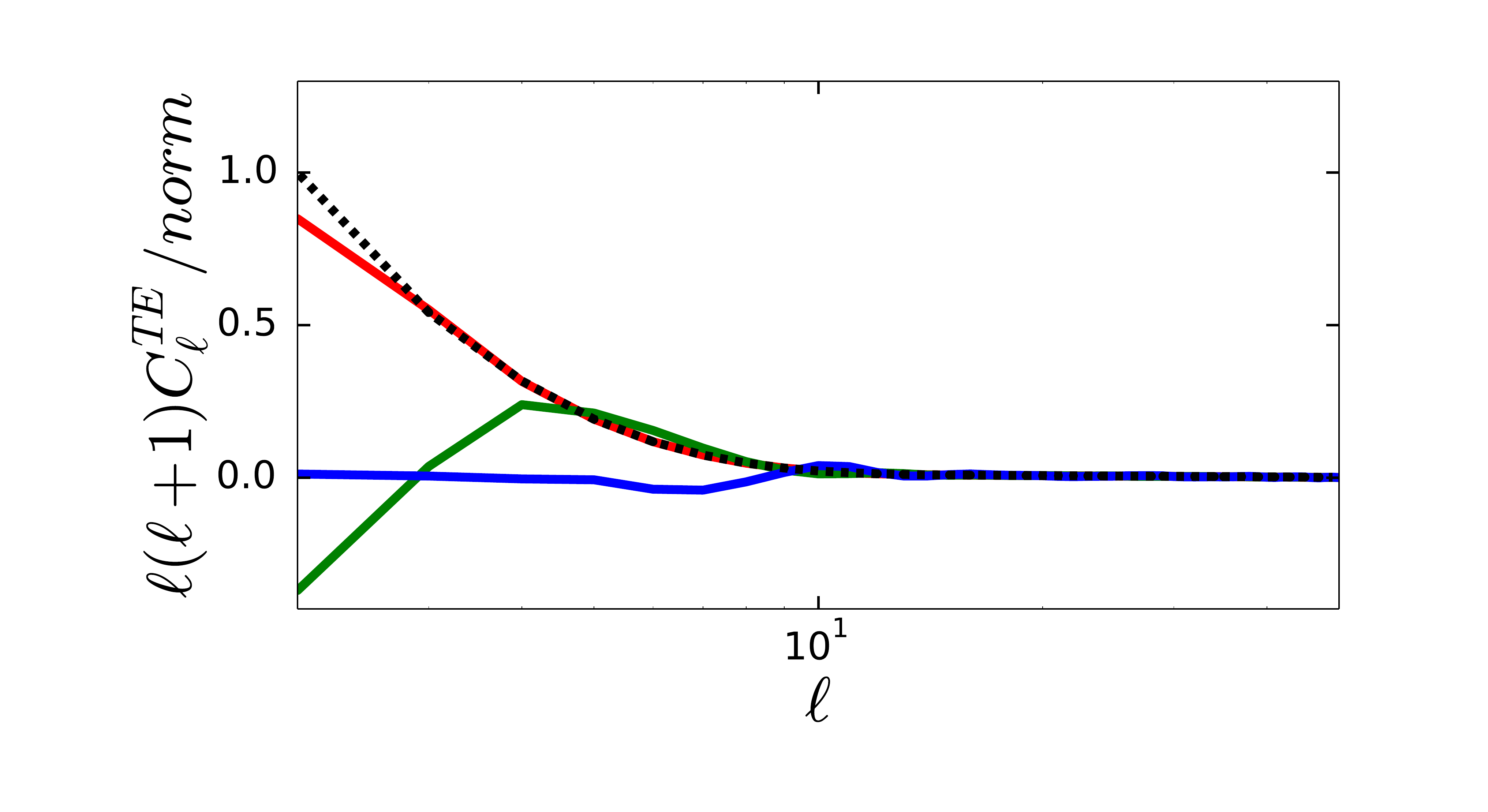}
\caption{$ET$ angular power spectrum for 3-tori of size $L=0.25D_{
\mathrm{LSS}}$ (blue), $L=0.5D_{
\mathrm{LSS}}$ (green), $L=D_{
\mathrm{LSS}}$ (red) and for the Euclidean space (dotted black line). The computation was done by taking into account the full transfer functions from CAMB. The norm is taken equal to $6C_{2}^{ET,\mathrm{iso}}$.}
\label{cl_ET}
\end{figure}

The cross-correlations between $B$ and $E$, or $B$ and $T$, disappear because of parity properties. There are only $EE$, $BB$, $ET$ and of course $TT$ correlations (already studied in section II.C.3.). In contrast to the temperature anisotropies, polarization is entirely induced by scattering at the LSS, and therefore cannot be present on scales much larger than the Hubble scale at the epoch of recombination. Polarized power spectra thus sharply decrease at low $\ell$, except for the contribution,on very large scales, of scattered radiation due to reionization~\cite{polarization}. 

Fig.~\ref{cl_EE} and \ref{cl_ET} show the the $EE$ and $ET$ power spectra respectively for different size of tori. As for $TT$ power spectra, there is a remarkable suppression of power on the largest scales for small tori, with oscillations at intermediate scales.

\subsection{Kullback-Leibler divergence and polarization}

The goal of this section is to use the polarization of the CMB with the Kullback-Leibler divergence.
The implementation of the Kullback-Leibler divergence is similar to what was done in Sec. III. We just have to change the correlation matrix.

\begin{figure}[!h]
\includegraphics[width=\columnwidth]{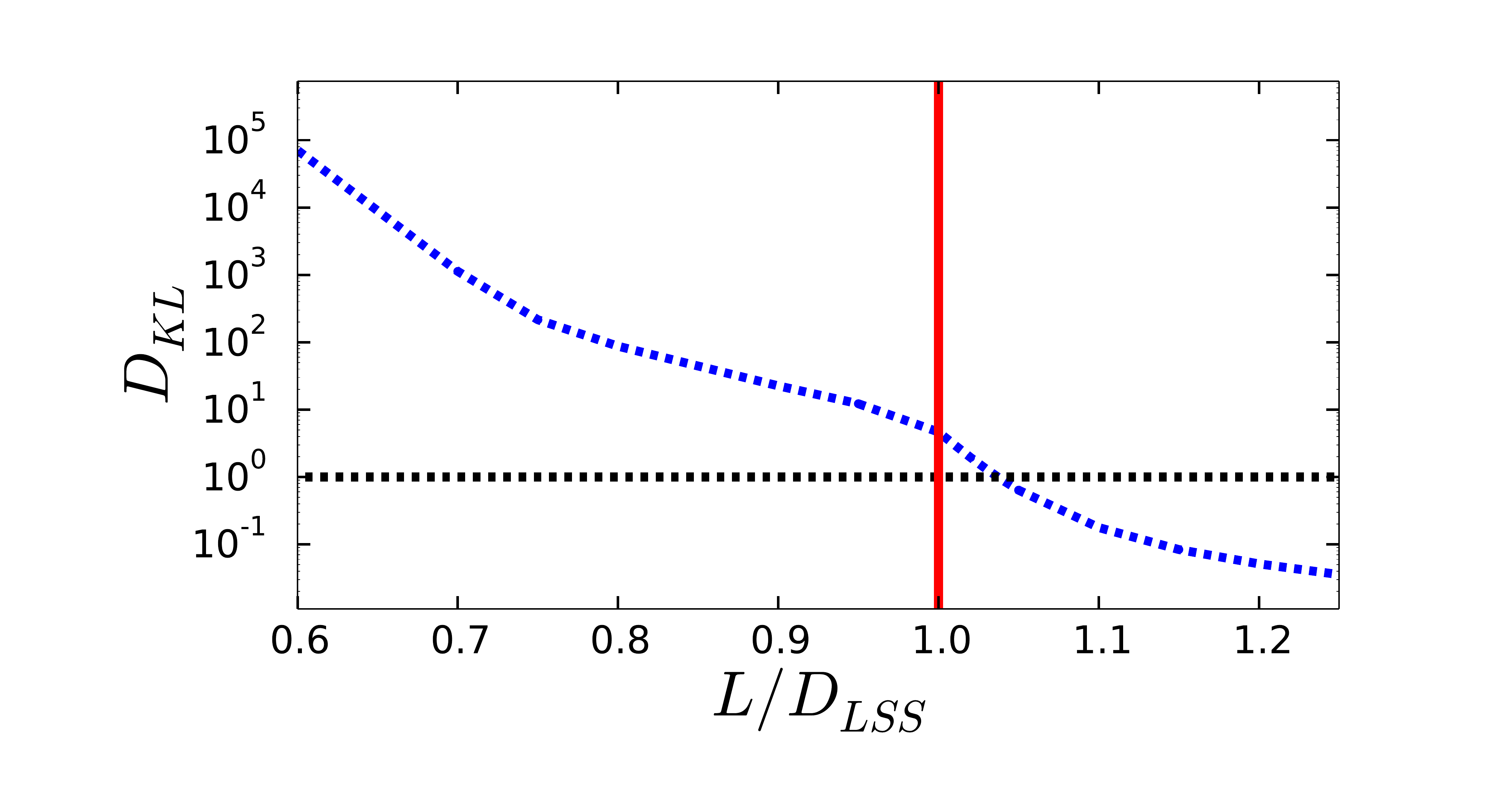}
\caption{Kullback divergence with $EE$ covariance matrices for ${\ell}_{\rm max}=20$, ${\ell}_{\rm cut}=20$ (blue line) as a function of the size of the cubic 3-torus.}
\label{dkl_EE}
\end{figure}

Fig.~\ref{dkl_EE} shows the Kullback-Leibler divergence of 3-tori models as a function of their size, based on $E$ mode polarization only. 
One can notice that the shape of the curve is smoother and less affected by the transition at $L=D_{
\mathrm{LSS}}$ than with pure temperature data $TT$. There is also an other transition 
at $L=0.6 D_{
\mathrm{LSS}}$ in addition to the transition at $L=D_{
\mathrm{LSS}}$. We can note that small tori are much better distinguished with $EE$ data than with $TT$ data. Unfortunately 
big tori are less constrained. The threshold of detection gives a boundary of $L_*=1.03D_{
\mathrm{LSS}}$ for the biggest cubic three-torus distinguishable with only $EE$ data in the ideal case with no noise and no mask.

These results are in very good agreement with~\cite{polarization-topology}. In these references, the "circles-in-the-sky" method is applied on simulated polarization data and it appears that polarization is better than temperature to perform the search of pairs of correlated circles into CMB data, and thus better to detect tori of size smaller than the diameter of the last scattering surface.

\subsection{Full covariance matrix}

One could be tempted to perform this analysis one more time with $ET$ matrices, but the matrices involved here being non-Hermitian, it is not possible. However, we consider the full (temperature,polarization) block correlation matrix

\begin{eqnarray}
C^{(2)}=\left[
\begin{array}{c|c}
C^{TT,\mathrm{torus}} & C^{TE,\mathrm{torus}} \\ \hline
C^{ET,\mathrm{torus}} & C^{EE,\mathrm{torus}}
\end{array}\right]
\end{eqnarray}
 
\begin{eqnarray}
C^{(1)}=\left[
\begin{array}{c|c}
C_{\ell}^{TT,\mathrm{iso}} & C_{\ell}^{TE,\mathrm{iso}} \\ \hline
C_{\ell}^{ET,\mathrm{iso}} & C_{\ell}^{EE,\mathrm{iso}}
\end{array}\right]
\end{eqnarray}
and compute the $D_{KL}$ results. $C^{(1)}$ is block diagonal and both $C^{(1)}$ and $C^{(2)}$ are still Hermitian. We will truncate each of the blocks of the matrices at $\ell_{cut}\leqslant \ell_{max}$ as before.

This generalization allows us to determine the best we can do with full temperature and polarization cosmic microwave background 
data in the ideal case.

The computation of formula Eq.~(\ref{formula_kullback}) needs to be generalized to the full covariance matrix

\begin{eqnarray}
 2D_{\mathrm{KL}}(1 || 2) = \ln \left[\frac{\det C^{(2)}}{\det C^{(1)}}\right]  +  \mathrm{trace}(M^T({C^{(2)}}^{-1})M)\nonumber\\
- 2[ \ell_{\rm cut}(\ell_{\rm cut}+2) - 3]
\end{eqnarray}
where $C^{(1)}$ is decomposed via a Cholesky decomposition with $M$ a lower triangular matrix

\begin{eqnarray}
C^{(1)}=M . M^T
\end{eqnarray} 

\begin{figure}[!h]
\includegraphics[width=\columnwidth]{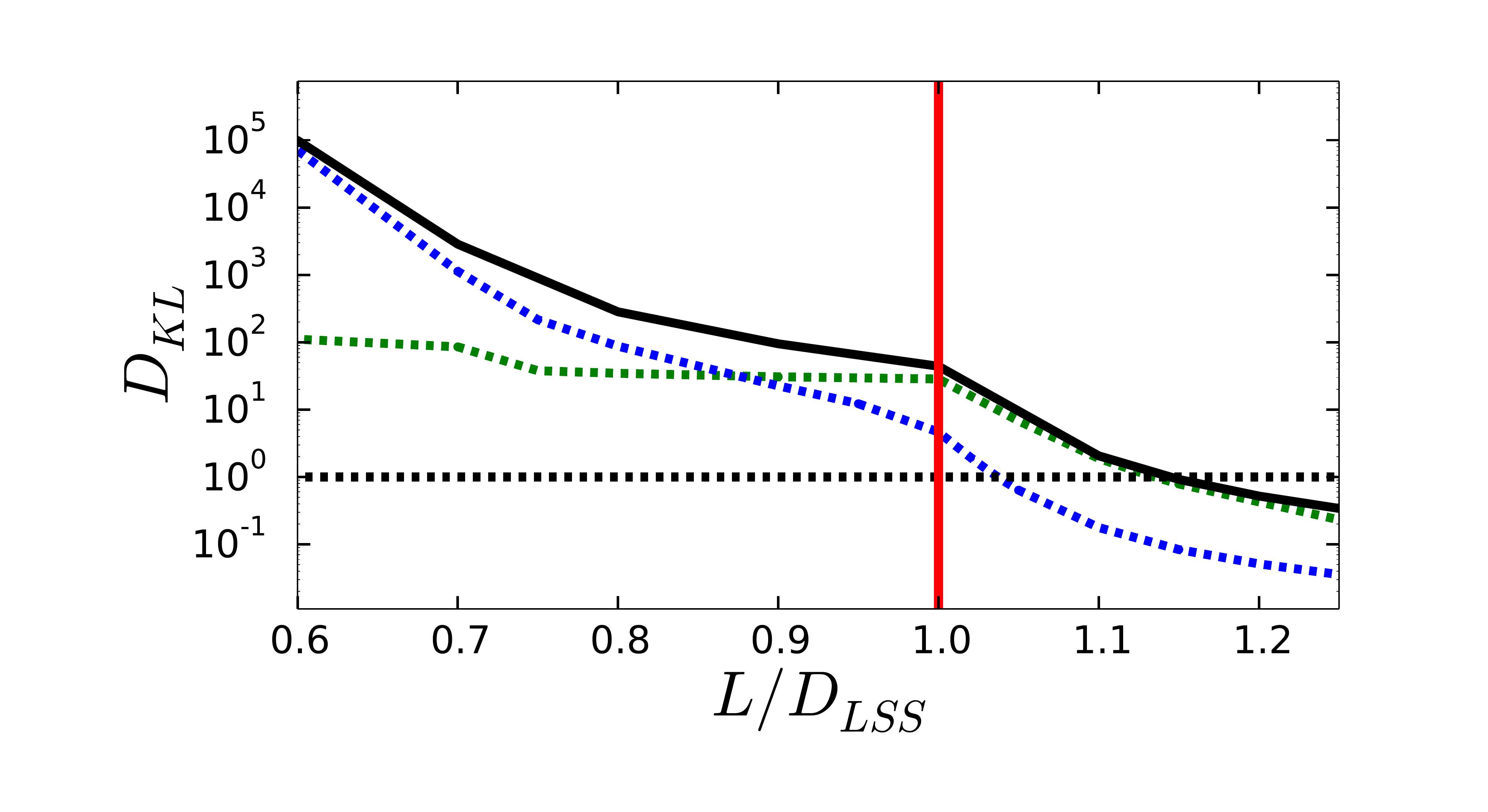}
\caption{$D_{\mathrm{KL}}$ with full covariance matrices (solid black line), $TT$ matrices only (dotted green line) and $EE$ matrices only (dotted blue line) at $\ell_{max}=20$} 
\label{dkl_full}
\end{figure}

Fig.~\ref{dkl_full} shows the evolution of $D^{\rm tot}_{\mathrm{KL}}$ (with the full covariance matrix) as a function of the size of the cubic 3-torus. 
Taking into account the full covariance matrix does not improve the detectability of cubic three-tori from 
Euclidean space at $L>D_{
\mathrm{LSS}}$. However around the size $L=0.86D_{
\mathrm{LSS}}$ where $D_{\mathrm{KL}}^{EE}$ and $D_{\mathrm{KL}}^{TT}$ curves cross there is a good improvement of the detectability due to the contribution of the polarization. In a nut shell we can say that for $L>D_{
\mathrm{LSS}}$ the detectability is determined by the temperature data, for small tori, by the polarization around $L=0.86D_{
\mathrm{LSS}}$ both temperature and polarization data should be taken into account.

Unfortunately, it means that if we are living in a universe bigger than the diameter of the last scattering surface $D_{
\mathrm{LSS}}$, the addition of the polarization will not improve significantly the constraints on the cosmic topology obtained with temperature data only.

\section{Conclusion}

This article has revisited the signatures of a non-trivial spatial topology on CMB anisotropy. After checking that our code recovers the standard results on the 2-point angular correlation function, we have focused on the correlation matrix. While the circles in the sky method allows one to efficiently probe topology in a model-independent way on scales smaller than the last scattering surface,we have focused on the isotropisation properties of the correlation matrix when the size of the fundamental polyhedron increases. This has been implemented in terms of the Kullback-Leiber distance.

Applied to the family of cubic tori, we have concluded that its size needs to be larger than 1.15 times the diameter of the last scattering surface in order for the finite space to be indistinguishable in practice from an infinite universe. The effect of the noise was considered and shown to be negligible for experiments such as WMAP and \emph{Planck} and we have also shown that for the conclusion is not affected by reasonable galactic cut.
We also applied this analysis to polarization simulation and discovered that when considering only polarization data, we have a smaller limit size of 
$L_*=1.03D_{
\mathrm{LSS}}$ for ideal experimental conditions. However we learned the polarization is more effective to distinguish small 3-tori. As a consequence, when both temperature and polarization data are taken into account, it improves the detectability only at $L<D_{
\mathrm{LSS}}$.

To finish, we have investigated the signature of a non-trivial topology on the 3-point function, focusing on the equilateral bispectrum for simplicity.

\emph{Planck} results on topology in~\cite{Planck_topology} show no detection of a non-trivial topology. On the one hand the \emph{Planck} collaboration find no evidence for the existence of back-to-back circles of correlation. The lower bound of any spatial dimension $L$ of the fundamental domain of our universe evaluated with this particular method is bigger than $0.94D_{
\mathrm{LSS}}$ with a confidence level of $99\%$. On the other hand the Bayesian analysis of the data show no strong evidence of a multi-connected universe, even if there is a faint detection of a torus bigger than the diameter of the last scattering surface. The future release of \emph{Planck} polarization data will allow us to improve considerably the constraints on models whose size $L$ is smaller than $D_{
\mathrm{LSS}}$ as explained in this paper. However, there is very few hope for an improvement of the constraints for models bigger than $D_{
\mathrm{LSS}}$ as seen earlier.

\begin{acknowledgments}
The authors would like to thank Alain Riazuelo for fruitful discussions and S. Rouberol for running the {\tt horizon} cluster where the
computations were carried. O.F. would like to thank J.Lef{\`e}vre for his help. This work made in the ILP LABEX (under reference ANR-10-LABX-63) was supported by French state funds managed by the ANR within the Investissements d'Avenir programme under reference ANR-11-IDEX-0004-02.
\end{acknowledgments}

\appendix
\begin{widetext}
\section{Spherical harmonics}\label{spherical_harmonics}

We summarize some basic formulas on spherical harmonics that have been used in this work. We refer to Ref.~\cite{vmk} for further properties.

The integral over three spherical harmonics can be obtained as
\begin{equation}\label{a1}
\int \dd^2 \hat{\bm n}\, Y_{{\ell}_{1} m_1}(\hat{\bm n})Y_{{\ell}_{2} m_2}(\hat{\bm n})Y_{{\ell}_{3} m_3}(\hat{\bm n})
  = \sqrt{\frac{(2{\ell}_{1}+1)(2{\ell}_{2}+1)(2{\ell}_{3}+1)}{4 \pi}}{\left( \begin{array}{ccc}
{\ell}_{1} & {\ell}_{2} & {\ell}_{3} \\
m_1 & m_2 & m_3 \\
\end{array} \right)}{\left( \begin{array}{ccc}
{\ell}_{1} & {\ell}_{2} & {\ell}_{3} \\
0 & 0 & 0 \\
\end{array} \right)}
\end{equation}
where the quantities on the r.h.s. are the Wigner $3j$-symbol. The first one, with no $m$-dependence, is non-vanishing only if $\ell_1+\ell_2+\ell_3$ is even.

The addition theorem is given by
\begin{equation}\label{a2}
 \sum_{m=-\ell}^{\ell}Y_{\ell m}( \hat{\bm n})Y_{\ell m}^{*}( \hat{\bm n})=\frac{2\ell+1}{4\pi}P_\ell( \hat{\bm n}. \hat{\bm n}').
 \end{equation}

If ${\ell}_{1}+{\ell}_{2}+{\ell}_{3}=2g$ with $g\in{\mathbb N}$, then the $3j$-symbol takes the form
\begin{equation}\label{a3}
\left( \begin{array}{ccc}
{\ell}_{1} & {\ell}_{2} & {\ell}_{3} \\
0 & 0 & 0 \\
\end{array} \right)=(-1)^{g}\sqrt{\frac{(2g-2{\ell}_{1})!(2g-2{\ell}_{2})!(2g-2{\ell}_{3})!}{(2g+1)!}}\frac{g!}{(g-{\ell}_{1})!(g-{\ell}_{2})!(g-{\ell}_{3})!}.
\end{equation}
\end{widetext}

\section{Non-Gaussianity}\label{section6}

The effect of the topology is to kill some wave-numbers and thus project the allowed perturbations in Fourier space to a subspace compatible with the boundary conditions imposed by the topology. As we have discussed in \S~\ref{section2}, the relation between the eigenmodes is linear so that this does not affect the statistical properties of the $a_{\ell m}$. Nevertheless, and as seen on Fig.~\ref{fig1b}, the topology has an imprint on the spectra.

The goal of this section is to generalize our former analysis of the 2-point correlation function to the 3-point function and understand the imprint of the topology on the angular bispectrum 
\begin{equation}\label{def_Blll}
  b^{m_1 m_2 m_3}_{{\ell}_{1} {\ell}_{2} {\ell}_{3}}=\langle a_{{\ell}_{1} m_1}a_{{\ell}_{2} m_2}a_{{\ell}_{3} m_3}\rangle.
\end{equation} 

\subsection{General formalism}

For a Gaussian temperature, $b^{m_1 m_2 m_3}_{{\ell}_{1} {\ell}_{2} {\ell}_{3}}=0$. Deviations from Gaussianity are expected to be due to the non-linear evolution of the perturbation~\cite{ngCP} or primordial non-Gaussianity generated during inflation~\cite{reviewNG}. 

From the 2-point correlation matrix, one can construct the angular power spectrum. Because of the violation of global isotropy, the 3-point function must be described by the 6-dimensional quantity $b^{m_1 m_2 m_3}_{{\ell}_{1} {\ell}_{2} {\ell}_{3}}$. As a first insight we however concentrate on the angular averaged bispectrum $B_{{\ell}_{1} {\ell}_{2} {\ell}_{3}}$, which is the analog of the angular power spectrum $C_\ell$ and is defined as 
\begin{equation}
B_{{\ell}_{1} {\ell}_{2} {\ell}_{3}}=\sum_{m_1 m_2 m_3}
{\left(\begin{array}{ccc}{\ell}_{1} & {\ell}_{2} & {\ell}_{3} \\
m_1 & m_2 & m_3 \\
\end{array} \right)}b^{m_1 m_2 m_3}_{{\ell}_{1} {\ell}_{2} {\ell}_{3}}.
\end{equation}
where ${\left(\begin{array}{ccc}
{\ell}_{1} & {\ell}_{2} & {\ell}_{3} \\
m_1 & m_2 & m_3 \\
\end{array} \right)}$ are the Wigner $3j$-symbols.

We described non-Gaussianity by decomposing the $a_{\ell m}$ as the sum of a Gaussian contribution $a_{\ell m}^{(\rm L)}$ and of a non-Gaussian one $a_{\ell m}^{(\rm NL)}$ as
\begin{equation}
a_{\ell m}=a_{\ell m}^{(\rm L)}+a_{\ell m}^{(\rm NL)}.
\end{equation}
In the standard description~\cite{non-gaussianities} with Euclidean trivial spatial topology, the $a_{\ell m}$ are given by
\begin{equation}
a_{\ell m}^{(\rm L)}={i}^{\ell}\int\frac{d^3\bm{k}}{{(2\pi)}^{3/2}}\phi_{\rm L}(\bm{k})G_\ell(k)Y_{\ell m}^*(\hat{\bm{k}}),
\end{equation}
which has to be compared to Eq.~(\ref{p4_2}). Similarly, the non-Gaussian contribution can be decomposed as
\begin{equation}
a_{\ell m}^{(\rm NL)}={i}^{\ell}\int\frac{d^3\bm{k}}{{(2\pi)}^{3/2}}\phi_{_{\rm NL}}(\bm{k})G_\ell(k)Y_{\ell m}^*(\hat{\bm{k}}).
\end{equation}
$\phi_{\rm L}(\bm{k})$ and $\phi_{_{\rm NL}}(\bm{k})$ correspond respectively of the Gaussian and non-Gaussian primordial metric perturbations and $G_\ell(k)$ the radiation transfer function described by Eq.~(\ref{swdopisw}). As usual, the 2-point correlation function of $\phi_{\rm L}$ defines the linear power spectrum as
\begin{equation}
 \langle\phi_{\rm L}({\bm k}_1)\phi_{\rm L}({\bm k}_2)\rangle ={(2\pi)}^{3} {\delta}^{(3)}(\bm{k}_1+\bm{k}_2)P_{\phi}(k_1).
\end{equation}
$\phi_{_{\rm NL}}$ is conveniently described by a function $f_{_{\rm NL}}$ defined from the 3-point function as
\begin{eqnarray}
\langle\phi_{\rm L}(\bm{k}_1)\phi_{\rm L}(\bm{k}_2)\phi_{_{\rm NL}}(\bm{k}_3)\rangle& =& 2{(2\pi)}^{3}f_{_{\rm NL}}(\bm{k}_1,\bm{k}_2,\bm{k}_3) P_{\phi}(k_1)\nonumber\\
 &&P_{\phi}(k_2){\delta}^{(3)}(\bm{k}_1+\bm{k}_2+\bm{k}_3)\nonumber\\
\end{eqnarray}
$f_{_{\rm NL}}$ is a function of the wave-numbers, the explicit form of which depends on the details of the inflationary model.

Let us now consider the case of a 3-torus. From the general expression~(\ref{p44t2}) and the particular expression of the coefficients $\xi$ given by Eq.~(\ref{xi-torus}) for a 3-torus, we have that the previous expressions now take the form
\begin{equation}
  a_{\ell m}^{(\mathrm{L})}=\frac{{(2\pi)}^{3}{i}^{\ell}}{V}\sum_{\bm{k}}\frac{1}{{(2\pi)}^{3/2}}\phi_{\rm L}(\bm{k})G_\ell(k)Y_{\ell m}^*(\hat{\bm k})
\end{equation}
and
\begin{equation}
a_{\ell m}^{(\mathrm{NL})}=\frac{{(2\pi)}^{3}{i}^{\ell}}{V}\sum_{\bm{k}}\frac{1}{{(2\pi)}^{3/2}}\phi_{_{\rm NL}}(\bm{k})G_\ell(k)Y_{\ell m}^*(\hat{\bm k})
\end{equation}
keeping in mind that the sum is taken on the wave-numbers defined by Eq.~(\ref{k-discret}). The 2-point and 3-point correlation functions are now defined as
\begin{equation}
 \langle\phi_{\rm L}(\bm{k}_1)\phi^*_{\rm L}(\bm{k}_2)\rangle = VP_{\phi}(k_1){\delta}_{\bm{k}_1,\bm{k}_2}
\end{equation}
and
\begin{equation}\label{deffnl}
\langle\phi_{\rm L}(\bm{k}_1)\phi_{\rm L}(\bm{k}_2)\phi_{_{\rm NL}}^*(\bm{k}_3)\rangle = 2Vf_{_{\rm NL}}P_{\phi}(k_1)P_{\phi}(k_2){\delta}_{\bm{k}_1+\bm{k}_2,\bm{k}_3}
\end{equation}
As $a_{{\ell}_{1} m_1}^{(\mathrm{L})}$ is Gaussian, we obviously have $\langle a_{{\ell}_{1} m_1}^{(\mathrm{L})}a_{{\ell}_{2} m_2}^{(\mathrm{L})}a_{{\ell}_{3} m_3}^{(\mathrm{L})}\rangle=0$ so that
\begin{eqnarray}
\langle a_{{\ell}_{1} m_1}a_{{\ell}_{2} m_2}a_{{\ell}_{3} m_3}\rangle&=& \langle a_{{\ell}_{1} m_1}^{(\mathrm{NL})}a_{{\ell}_{2} m_2}^{(\mathrm{L})}a_{{\ell}_{3} m_3}^{(\mathrm{L})}\rangle+ ({\rm perm.})\nonumber\\
  &&\qquad +\mathcal{O}({f_{_{\rm NL}}}^2).
\end{eqnarray}.

\subsection{Bispectrum in a 3-torus}

From the previous definitions, it is easily checked that for a 3-torus
\begin{widetext}
\begin{eqnarray}
\langle a_{{\ell}_{1} m_1}^{(\mathrm{L})}a_{{\ell}_{2} m_2}^{(\mathrm{L})}a_{{\ell}_{3} m_3}^{(\mathrm{NL})}\rangle
 &=& \frac{{(2\pi)}^{9/2}{i}^{{\ell}_{1}+{\ell}_{2}+{\ell}_{3}}}{V^3}
  \sum_{{\bm k}_1,{\bm k}_2,{\bm k}_3}\langle\phi_{\rm L}({\bm k}_1)\phi_{\rm L}({\bm k}_2)\phi_{_{\rm NL}}({\bm k}_3)\rangle
G_{\ell_1}(k_1)G_{\ell_2}(k_2)G_{\ell_3}(k_3) \nonumber \\
 &&\qquad\qquad\qquad\qquad\qquad\qquad Y_{{\ell}_{1} m_1}^*(\hat{\bm k}_{1})Y_{{\ell}_{2} m_2}^*(\hat{\bm k}_{2})Y_{{\ell}_{3} m_3}^*(\hat{\bm k}_{3}).
\end{eqnarray}
Using the definition~(\ref{deffnl}) and then summing on ${\bm k}_3$, this reduces to
\begin{eqnarray}
\langle a_{{\ell}_{1} m_1}^{(\mathrm{L})}a_{{\ell}_{2} m_2}^{(\mathrm{L})}a_{{\ell}_{3} m_3}^{(\mathrm{NL})}\rangle&=&\frac{2{(2\pi)}^{9/2}{i}^{{\ell}_{1}+{\ell}_{2}+{\ell}_{3}}}{V^2}\sum_{{\bm k}_1,{\bm k}_2}f_{_{\rm NL}}({\bm k}_1,{\bm k}_2)
P_{\phi}(k_1)P_{\phi}(k_2)G_\ell(k_1)G_\ell(k_2)G_\ell(|{\bm k}_1+{\bm k}_2|)\nonumber\\
&&\qquad\qquad\qquad\qquad\qquad\qquad Y_{{\ell}_{1} m_1}^*(\hat{\bm k}_{1})Y_{{\ell}_{2} m_2}^*(\hat{\bm k}_{2})Y_{{\ell}_{3} m_3}^*(-\hat{\bm k}_{12}),
\end{eqnarray}
where we have defined $\hat{\bm k}_{12}\equiv({\bm k}_1+{\bm k}_2)/|{\bm k}_1+{\bm k}_2|$. The bispectrum $B_{\ell_1\ell_2\ell_3}$ is obtained by contracting with the Wigner $3j$-symbol and summing on $(m_1,m_2,m_3)$. Thanks to the property~(\ref{a1}), we can replace the $3j$-symbol by an integral over three spherical harmonics to get
\begin{eqnarray}
B_{{\ell}_{1} {\ell}_{2} {\ell}_{3}} &=& \frac{6{(2\pi)}^{9/2} i^{\ell_1+\ell_2+\ell_3}}{V^2\left( \begin{array}{ccc}
{\ell}_{1} & {\ell}_{2} & {\ell}_{3} \\
0 & 0 & 0 \\
\end{array} \right)}   
\sqrt{\frac{4 \pi}{(2{\ell}_{1}+1)(2{\ell}_{2}+1)(2{\ell}_{3}+1)}}
\sum_{{\bm k}_1,{\bm k}_2}   f_{_{\rm NL}}({\bm k}_1,{\bm k}_2)  P_{\phi}(k_1)P_{\phi}(k_2)G_{{\ell}_{1}}(k_1)G_{{\ell}_{2}}(k_2)
G_{{\ell}_{3}}(|{\bm k}_1+{\bm k}_2|)  \nonumber\\
&&
\int \dd^2\hat{\bm n}\left[\sum_{m_1}Y_{{\ell}_{1} m_1}(\hat{\bm n})Y_{{\ell}_{1} m_1}^*(\hat{\bm k}_{1})\right]
\left[\sum_{m_2}Y_{{\ell}_{2} m_2}(\hat{\bm n})Y_{{\ell}_{2} m_2}^*(\hat{\bm k}_{2})\right]
\left[\sum_{m_3}Y_{{\ell}_{3} m_3}(\hat{\bm n})Y_{{\ell}_{3} m_3}^*(-\hat{\bm k}_{12})\right].
\end{eqnarray}
We remind that ${\ell}_{1}+{\ell}_{2}+{\ell}_{3}$ is even. Each sum over $m_i$ with $i\in \{1,2,3\}$ can be expressed in terms of Legendre polynomials using Eq.~(\ref{a2}) so that the bispectrum takes the simple form
\begin{eqnarray}\label{xbb}
B_{{\ell}_{1} {\ell}_{2} {\ell}_{3}}= \frac{\beta_{\ell_1\ell_2\ell_3}}{V^2} 
\sum_{{\bm k}_1,{\bm k}_2} f_{_{\rm NL}}({\bm k}_1,{\bm k}_2)  P_{\phi}(k_1)P_{\phi}(k_2)G_{{\ell}_{1}}(k_1)G_{{\ell}_{2}}(k_2)
G_{{\ell}_{3}}(|{\bm k}_1+{\bm k}_2|)I_\ell({\bm k}_1,{\bm k}_2)
\end{eqnarray}
\end{widetext}
with
\begin{equation}
I_\ell({\bm k}_1,{\bm k}_2) \equiv \int \dd^2 \hat{\bm n}
    P_{\ell}(\hat{\bm n}.\hat{\bm k}_{1})
    P_{\ell}(\hat{\bm n}.\hat{\bm k}_{2})
    P_{\ell}\left(\hat{\bm n}.\hat{\bm k}_{12}\right)
\end{equation}
and
\begin{equation}
 \beta_{\ell_1\ell_2\ell_3} = 3\sqrt{2}{\pi}^2{i}^{{\ell}_{1}+{\ell}_{2}+{\ell}_{3}}
   \frac{\sqrt{(2{\ell}_{1}+1)(2{\ell}_{2}+1)(2{\ell}_{3}+1)}}{{\left( \begin{array}{ccc}
{\ell}_{1} & {\ell}_{2} & {\ell}_{3} \\
0 & 0 & 0 \\
\end{array} \right)}}.
\end{equation}
When ${\ell}_{1}={\ell}_{2}={\ell}_{3}=\ell$ and ${\ell}_{1}+{\ell}_{2}+{\ell}_{3}$ is even, this coefficient reduces to 
\begin{equation}
 \beta_{\ell \ell \ell}= 3{\pi}^2 \sqrt{2 (2\ell+1)^3}
\sqrt{\frac{(3\ell+1)!}{{(\ell!)}^3}}\frac{{(\frac{\ell}{2})!}^3}{(\frac{3\ell}{2})!}.
\end{equation}
The equation~(\ref{xbb}) is general and can be computed as soon as $f_{_{\rm NL}}({\bm k}_1,{\bm k}_2)$ and $P_\phi(k)$ are known.

\subsection{Computation of $I_\ell({\bm k}_1,{\bm k}_2) $}

The previous expressions can be further simplified since the kernel $I_\ell({\bm k}_1,{\bm k}_2) $ can be computed analytically.
First, defining ${\mu}_{12}$ and $K$ as
\begin{equation}
   {\mu}_{12}\equiv \hat{\bm k}_{1}.\hat{\bm k}_{2}=\cos({\beta}_{12}), \qquad
   K=\frac{k_1}{k_2},
\end{equation}
$I_\ell({\bm k}_1,{\bm k}_2) $ can be written as as function of $(K,\mu_{12})$. To that purpose, we define
${\bm k}_{+}$, ${\bm k}_{-}$ and $\hat{\bm u}$ as
\begin{eqnarray}
{\bm k}_{+}&=&\hat{\bm k}_{2}+\hat{\bm k}_{1}, \\
{\bm k}_{-}&=&\hat{\bm k}_{2}-\hat{\bm k}_{1}, \\
\hat{\bm u}&=&\hat{\bm k}_{+}\wedge\hat{\bm k}_{-}.
\end{eqnarray}
They clearly satisfy ${\bm k}_{+}.{\bm k}_{-}=0$. Now, any vector ${\bm n}$ can be decomposed as
\begin{equation}
 {\bm n}= {\bm n}_\parallel + n_\perp\hat{\bm u}
\end{equation}
where the first term is the projection on the plane defined by $(\hat{\bm k}_{1},\hat{\bm k}_{2})$ and ${\bm n}_\perp=n_\perp\hat{\bm u}$ is perpendicular to this plane.

With $\theta$ being the angle between $\bm{n}$ and $\hat{\bm u}$, one has $n_{\perp}=\cos{\theta}$ and $n_\parallel\sin{\theta}$. Then, introducing $\alpha$ the angle between $\hat{\bm n}$ and $\hat{\bm k}_+$, we have the relations
\begin{eqnarray}
 {\bm n}.{\bm k}_+&=&{\bm n}_\parallel {k}_+\cos\alpha\\
 {\bm n}.{\bm k}_-&=&n_\parallel {k}_-\sin\alpha \\
 \bm{n}.\hat{\bm k}_1&=&n_\parallel\frac{\cos\alpha-\sin\alpha}{2}\\
  \bm{n}.\hat{\bm k}_2&=&n_\parallel\frac{\cos\alpha+\sin\alpha}{2},
\end{eqnarray}
from which we deduce that
\begin{eqnarray}
\bm{n}.{\bm k}_{12} &=&  \frac{k_1(\cos\alpha-\sin\alpha)+k_2(\cos\alpha+\sin\alpha)}
{2\sqrt{{k_1}^2+{k_2}^2+2k_{1}k_{2}}},
\end{eqnarray}
which can also be rewritten as
\begin{eqnarray}
\bm{n}.{\bm k}_{12} &=&  n_\parallel\frac{(1+K)\cos\alpha+(1-K)\sin\alpha}
{\sqrt{1+2\mu_{12}K + K^2}}.
\end{eqnarray}
It follows that $I_\ell$ can be expressed only as a function of $(K,\mu_{12})$ after integration on $\theta$ and $\alpha$ as
\begin{widetext}
\begin{eqnarray}
I_\ell(K,{\mu}_{12})&=&\int \dd\theta \sin\theta\dd\alpha \nonumber \\
&& P_\ell\left(\frac{\sin\theta\left[\cos\alpha-\sin\alpha\right]}{2}\right)
      P_\ell\left(\frac{\sin\theta\left[\cos\alpha+\sin\alpha\right]}{2}\right)
      P_\ell\left(\sin\theta\frac{(1+K)\cos\alpha+(1-K)\sin\alpha}{\sqrt{1+2\mu_{12}K + K^2}} \right).
\end{eqnarray}
\end{widetext}
Given that the Legendre polynomials are of order $\ell$, this is an integration of a product of sine and cosine. We thus expect $I_{\ell}$ to be of the form
$$
I_{2p}(K,\mu)=\frac{\sum_{i=0}^{2p}\sum_{j=0,i+j \leqslant 2p}^{i}a^{(p)}_{ij}K^i{\mu}^j}{{(1 + K^2 + 2 K \mu)}^p}
$$ 
with $a^{(p)}_{i0}=0$ for $i$ odd, and $i+j \leqslant 2p$ with $j\leqslant i$. These coefficients can be easily computed by \emph{e.g.} a Mathematica code. As an example, the two first $I_{\ell}$ functions are given by
\begin{eqnarray}
 I_{0} &=& 4 \pi \nonumber\\
 I_{2}  &=& -\frac{\pi\left[\frac{43}{560}(K^2+1)+\frac{3}{20} \mu K\right]}{K^2+2 \mu K+1}.
\end{eqnarray}

In conclusion, the bispectrum reduces to the triple sum
\begin{eqnarray}\label{xbb1}
B_{{\ell}{\ell} {\ell}}&=& \frac{\beta_{\ell \ell \ell}}{V^2}  \sum_{K,k_2,\mu_{12}} f_{_{\rm NL}}(K,k_2,\mu_{12})  P_{\phi}(Kk_2)P_{\phi}(k_2)\nonumber\\
&&G_{{\ell}}(k_2)G_{{\ell}}(Kk_2)
G_{{\ell}}(k_2\sqrt{1+2\mu_{12}K+K^2})\nonumber\\
&& I_{\ell}(K,\mu_{12}).
\end{eqnarray}

Recent \emph{Planck} results in~\cite{Planck_NG} show no strong evidence of possible primordial non-gaussianities.


\end{document}